\documentclass[letterpaper,journal]{IEEEtran}
\usepackage{amsmath,amsfonts}
\usepackage{algorithmicx,algpseudocode}
\usepackage{algorithm}
\usepackage{array}
\usepackage[caption=false,font=normalsize,labelfont=sf,textfont=sf]{subfig}
\usepackage{textcomp}
\usepackage{stfloats}
\usepackage{url}
\usepackage{verbatim}
\usepackage{graphicx}
\usepackage{cite}

\usepackage{diagbox}
\usepackage{multirow} 
\usepackage{xcolor}
\begin{document}

\title{MSADM: Large Language Model (LLM) Assisted
End-to-End Network Health Management \\
Based on Multi-Scale Semanticization}

\author{Fengxiao Tang,~\IEEEmembership{Senior Member,~IEEE}, \textit{Xiaonan Wang,~\IEEEmembership{Graduate Student Member,~IEEE}}, \\Xun Yuan,~\IEEEmembership{Graduate Student Member,~IEEE}, Linfeng Luo,~\IEEEmembership{Graduate Student Member,~IEEE}, \\Ming Zhao,~\IEEEmembership{Member,~IEEE}, Tianchi Huang,~\IEEEmembership{Member,~IEEE} and Nei Kato,~\IEEEmembership{Fellow,~IEEE}
\thanks{This work was supported in part by the National Natural Science Foundation
of China under Grant 62302527, in part by Hunan Provincial Natural
Science Foundation under Grant 2023jj40774. (Corresponding author: Ming Zhao.)}
\thanks{
 Fengxiao Tang, Xiaonan Wang, Xun Yuan, Linfeng Luo, and Ming Zhao are with the School of Computer Science and Engineering, Central South University, Changsha 410083, China (e-mail: tangfengxiao@csu.edu.cn; wxiaonan068@gmail.com; yuan.xun@csu.edu.cn; 
 luolinfeng@csu.edu.cn;
 meanzhao@csu.edu.cn).
}

\thanks{
Tianchi Huang, Computer Science and Technology, Tsinghua University, Beijing, 100-084, China (e-mail: mythkastgod@gmail.com).
}
\thanks{
Fengxiao Tang, Xun Yuan, and Nei Kato are with the Graduate School of Information Sciences (GSIS),
Tohoku University, Sendai 980-8576, Japan (e-mail: kato@it.is.tohoku.ac.jp).
}
}




\maketitle

\begin{abstract}
Network device and system health management is the foundation of modern network operations and maintenance. Traditional health management methods, relying on expert identification or simple rule-based algorithms, struggle to cope with the heterogeneous networks (HNs) environment. Moreover, current state-of-the-art distributed fault diagnosis methods, which utilize specific machine learning techniques, lack multi-scale adaptivity for heterogeneous device information, resulting in unsatisfactory diagnostic accuracy for HNs. In this paper, we develop an LLM-assisted end-to-end intelligent network health management framework. The framework first proposes a multi-scale data scaling method based on unsupervised learning to address the multi-scale data problem in HNs. Secondly, we combine the semantic rule tree with the attention mechanism to propose a Multi-Scale Semanticized Anomaly Detection Model (MSADM) that generates network semantic information while detecting anomalies. Finally, we embed a chain-of-thought-based large-scale language model downstream to adaptively analyze the fault diagnosis results and create an analysis report containing detailed fault information and optimization strategies. We compare our scheme with other fault diagnosis models and demonstrate that it performs well on several metrics of network fault diagnosis.
\end{abstract}


\section{Introduction}
\IEEEPARstart{I}{n} recent years, owing to the rapid development of communication and unmanned control technologies, as well as the increasing demand for network connectivity, there has been a significant trend.  Heterogeneous networks (HNs) have thus become an integral part of people’s daily lives. According to the $the~World~Internet~Development~Report~2024$~\cite{ChineseAcademyofCyberspaceStudies2024}, by the end of $2024$, the global deployment of 5G base stations exceeded $3.64$ million, with the total number of 5G connected users surpassing $1.01$ billion, covering $33.1\%$ of the worldwide population. HNs integrate diverse communication devices, such as base stations, UAVs, and vehicles, across varied environments~\cite{11244851}. These networks are critical in domains like emergency communications, transportation, and military operations~\cite{THz,9130088,10644103}. To improve the availability and reliability of HNs, we must perform timely health management to detect network anomalies and diagnose network failures.\par
Recent methods of network health management (NHM) can be categorized into two types: Config-Based Detection (CBD) and General Feature Engineering-Based Anomaly Detection (GFE-AD). CBD identifies anomalies by monitoring wireless measurements and comparing them with predefined thresholds or rules~\cite{10981691,szilagyi2012automatic,khatib2015diagnosis,zou2024fta,10113753}. However, despite their efficiency, such methods require extensive manual labeling and rule maintenance, consume significant time and labor, and struggle to adapt to dynamically changing network environments. As a result, researchers have turned to machine learning methods to directly model network information. Among them, classification-based models directly output anomaly detection results through training~\cite{nti2022mini}, while diffusion-based methods learn the distribution of normal data and identify anomalies by calculating the difference between actual features and generated features~\cite{tuli2022tranad,10113753,chen2023imdiffusion}.\par
\begin{figure}
\centering
\includegraphics[width=1\columnwidth]{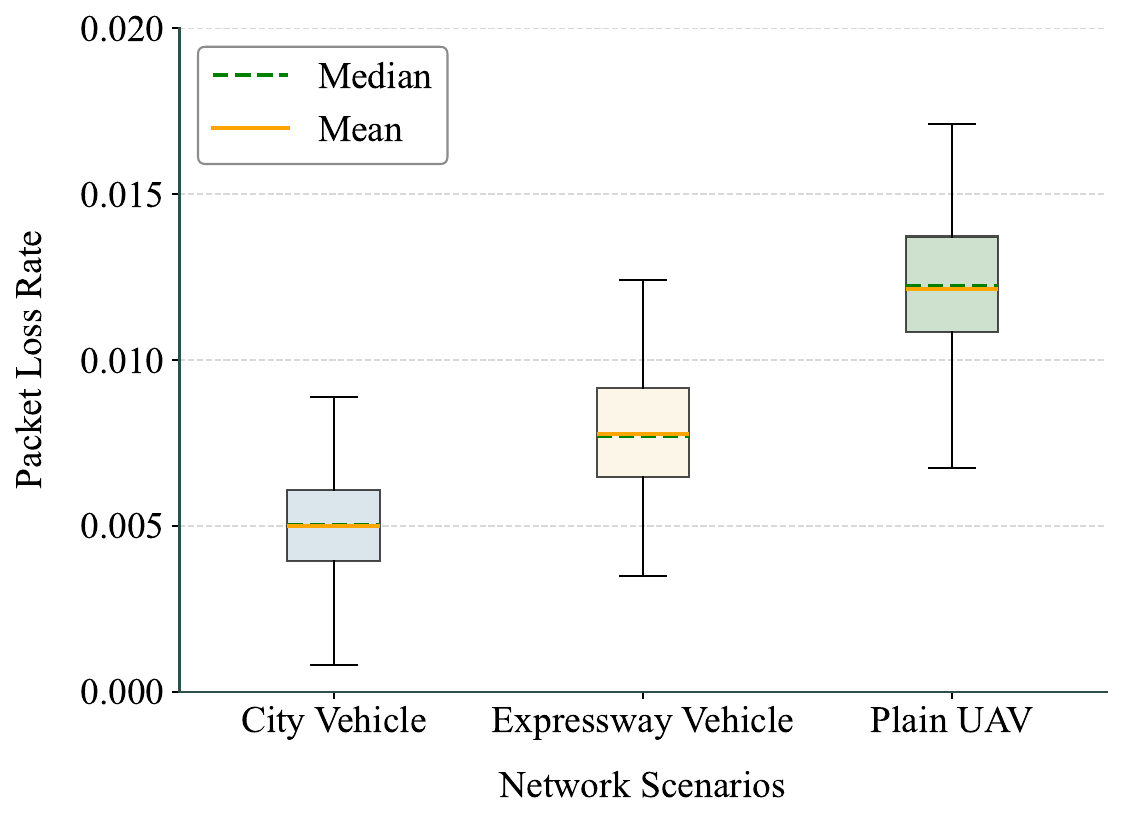}
\caption{  Packet loss rate intervals are shown for in-vehicle networks in cities and on highways, as well as for smooth-flying drone networks. The green dashed line represents the median, and the orange line represents the mean. }
\label{03mutidevice}
\end{figure}

Unfortunately, our analysis of the traffic information in existing HNs (As shown in Fig.~\ref{03mutidevice}, we present two types of communication nodes in the same environment and the same type of nodes in different environments to highlight the variability in anomaly manifestations.) reveals significant diversity in anomaly manifestations across different communication nodes. Such findings challenge the effectiveness of GFE-AD in HNs and emphasize the multi-scale nature of data representation in such environments.  \par

Therefore, we propose a Multi-Scale Semanticized Anomaly Detection Model (MSADM). Its two-stage process is designed to first perform network anomaly detection and fault diagnosis using a GFE-AD model, and then translate the unified severity levels into natural language descriptions via a semantic rule tree. These descriptions guide an LLM to generate comprehensive reports and mitigation solutions, enabling end-to-end health management from detection to mitigation.\par

Unlike previous GFE-AD methods~\cite{tuli2022tranad}, we propose a multi-scale normalization-based anomaly detection model for HNs~\cite{hayat2016survey}. Our key idea is to build a dynamic rule base that maps the multi-scale anomaly behaviors of heterogeneous communication nodes into unified anomaly severity levels. This mapping accounts for the fact that different nodes may have varying anomaly thresholds for the same performance metric. For instance, for one node, a delay of 100 milliseconds might be critical, but for another, it is normal. Based on this rule base, we can align these distinct thresholds into standardized levels, such as "low/medium/high". However, anomaly detection alone cannot address the fault mitigation phase in the NHM lifecycle.\par

Fortunately, LLM has demonstrated powerful text generation and complex reasoning capabilities in multiple fields, especially with research proving its outstanding performance in network semantic parsing~\cite{10.1145/3651890.3672268,wu2024nextgptanytoanymultimodalllm}.  We believe that LLM can also be applied to the fault mitigation phase for solution generation. To integrate Large Language Model (LLM) into network health management, we introduce the novel concept of Network Semanticization and have designed a corresponding scheme for our MSADM. This scheme uses a semantic rule tree to generate textual network status reports and guides the LLM to produce fault mitigation solutions. In the early phase of the scheme, the semantic rule tree translates the anomaly state list of performance metrics into natural language descriptions, enabling an intuitive representation of network information (Section~\ref{section:model}). In the later phase, these textual descriptions serve as part of the prompt to guide the LLM in comprehending network states and generating standardized network reports with mitigation suggestions (Section~\ref{section:llm}). \par
We evaluate MSADM against time-series anomaly detection schemes in HNs via real network data. Results show that the multi-scale normalization-based anomaly detection model achieves an average detection accuracy of $97.09\%$ on heterogeneous nodes, such as base stations, vehicle networks, outperforming existing schemes. For fault diagnosis, MSADM attains an average classification accuracy of $89.42\%$ on heterogeneous communication nodes, surpassing traditional rule engines, for example, TranAD~\cite{tuli2022tranad} by at least $22\%$, highlighting the critical role of data normalization in heterogeneous network health management. Further analysis reveals that compared to directly feeding raw data into LLM, the semantic rule tree-based network semanticization scheme significantly improves report readability, providing administrators with more precise mitigation guidance, for example, "Prioritize bandwidth expansion for Node B". \par
In general, we summarize the contributions as follows:\par

\begin{itemize}
\item {To address the inconsistent detection benchmarks across heterogeneous nodes, a multi-scale normalization mechanism based on a dynamic rule base is proposed. It unifies the quantification of performance metrics according to node-specific contexts, thereby resolving the issue of accuracy degradation in anomaly detection.}

\item {We introduce and design a "network semanticization" process. Utilizing a semantic rule tree, this process maps low-level anomaly states into structured natural language descriptions. This method provides textualized network information to LLM, enabling them to generate comprehensive diagnostic reports and propose mitigation suggestions.}

\item{ We integrate multi-scale normalization and network semanticization into a unified framework, MSADM, achieving end-to-end management from anomaly detection to fault mitigation. Experimental results demonstrate the effectiveness of this framework in enhancing overall performance for heterogeneous network health management tasks.}
\end{itemize}

\begin{figure*}[thb]
\centering
\includegraphics[width=1\textwidth]{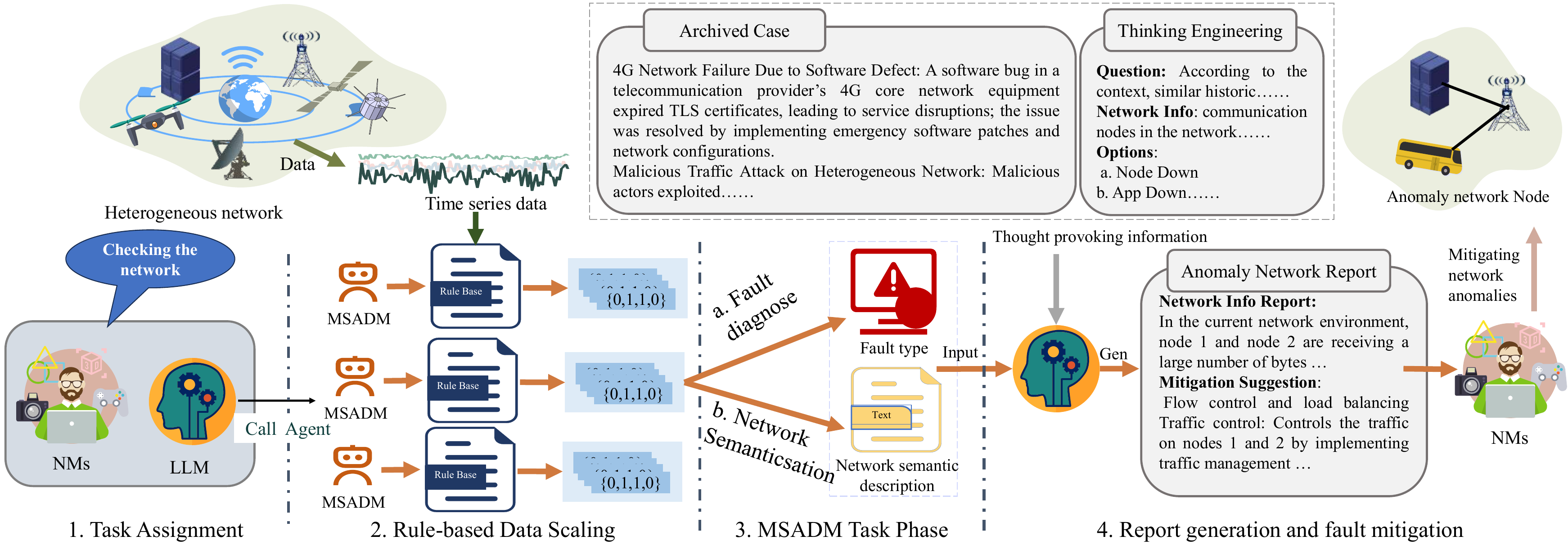}
\caption{Heterogeneous network health management scheme architecture: MSADM collects and normalizes network information to generate a status list. Subsequently, MSADM executes dual processing: it performs anomaly detection on the status list using a detection model based on multi-scale normalization while simultaneously transforming the list into natural language information via a semantic rule tree. Finally, both the anomaly detection results and network semanticization outputs are fed into an LLM. The LLM synthesizes these inputs to generate a comprehensive network analysis report and a corresponding fault mitigation solution.
}
\label{04model}
\end{figure*}

\section{Related Work}
\label{section:relatedwork}
Our health management scheme includes the detection of network anomalies, the semanticization of the network, and the mitigation of network faults. Since we propose network semanticization for the first time, the related work here focuses on anomaly detection and fault mitigation. \par

\subsection{Anomaly Detection } Traditional anomaly detection models identify anomalies by monitoring wireless measurements and comparing them to predefined criteria. For instance, researchers in~\cite{khanafer2008automated} developed criteria using simulated and actual data to enhance detection accuracy. Bayesian-based classification methods~\cite{barco2008continuous,barco2010learning,jin2015detecting,chen2024aigc} leverage probability and graph theory to systematically uncover the root causes of anomalies. However, these methods rely heavily on extensive historical anomaly data, and establishing criteria requires significant prior knowledge, labor, and time, limiting their scalability and practicality. \par
Machine learning techniques excel at mining potential information from data, enabling the rapid and accurate detection of subtle changes in network statuses and key performance indicators (KPIs). For example, TranAD~\cite{tuli2022tranad} uses a deep transformer-based model with an attention sequence encoder to detect anomalies using temporal trends. Similarly, Dcdetector~\cite{yang2023dcdetector} uses a dual attention mechanism and comparative learning to improve the representation of the anomaly sample. Despite these advancements, challenges remain in wireless networks. The scarcity of anomaly samples often leads to overfitting, and models trained for specific devices cannot generalize to nodes deployed in different environments. Consequently, multiple models are required to fit diverse network structures, significantly increasing training costs. \par
Existing distributed methods~\cite{boem2019distributed,qian2021detection} often treat anomalies as a unified phenomenon, overlooking the unique representations of individual communicating entities. This approach faces practical challenges, as it does not address the specific characteristics of anomalies across different network components.\par


\subsection{Fault Mitigation}
The study~\cite{10.1016/j.engappai.2004.08.031} addressed the limitations of traditional redundancy approaches, which are impractical for large-scale, real-time embedded systems due to budgetary and hardware restrictions, as well as the critical consequences of system failures. They achieved this by introducing a hierarchical autonomous fault mitigation framework rooted in model-integrated computation. Nevertheless, their framework encounters obstacles, including limited adaptability to unforeseen dynamic faults, inadequate assessment of the hierarchical architecture's scalability, constrained diagnostic decision-making intelligence, and the necessity to balance real-time performance with resource utilization in intricate environments like HNs, particularly concerning the trade-off between real-time performance and resource overhead.


In a separate investigation, the researchers addressed the issue of inefficient implicit fault localization stemming from the collaboration of multiprotocol devices within heterogeneous communication networks. They used a rule-based expert system (NDRBES) featuring hierarchical diagnostic tree logic. However, this system struggles with poor adaptability to evolving network topologies, lacks comprehensive cross-vendor protocol compatibility, and exhibits shortcomings in optimizing rules tailored for resource-constrained devices operating in HNs.


An alternative solution was devised to enhance link redundancy efficiency and reduce failover delays caused by the coexistence of multi-protocol devices in heterogeneous communication networks. This was accomplished by integrating the Rapid Spanning Tree Protocol (RSTP) with a dual-ring topology design. Despite these advancements, the solution still grapples with challenges such as inadequate dynamic topology responsiveness, persistent cross-vendor protocol interoperability issues, and difficulties in synchronizing delays across heterogeneous links within such networks.\par


\section{Framework}
\label{section:framework}
As illustrated in Fig.~\ref{04model}, MSADM is architected as a two-stage pipeline to enable end-to-end network health management. The entire process begins with raw time-series data from heterogeneous network nodes (e.g., UAVs, base stations) and culminates in actionable reports and mitigation solutions. The following subsections detail the four core components of our framework.

\textbf{Rule-based Multi-Scale Normalization:}
To achieve cross-device adaptability, MSADM first normalizes heterogeneous metrics into a unified scale. We design a dynamic rule library that is updated via unsupervised learning from historical device data. This library transforms each time-series feature into a multi-state abnormality list (e.g., `low,' `medium,' `high'), effectively creating a standardized representation of anomalous behaviors regardless of the source device's specific performance characteristics.

\textbf{Multi-Scale Anomaly Detection \& Diagnosis:}
The normalized state lists serve as the input to our dedicated anomaly detection model. This model is engineered to address the intricacies of network data by incorporating specialized attention mechanisms, including temporal and channel attention. These components allow the model to accurately capture performance dynamics and dependencies, thereby executing precise anomaly detection and fault diagnosis tasks.

\textbf{Network Information Semanticization:}
To bridge the gap between numerical data and LLM understanding, we introduce a semanticization module. This component translates the normalized state lists and anomaly results into structured natural language descriptions. The generated text explicitly delineates the relationships between node KPIs, their performance characteristics, and their multi-dimensional anomaly levels. This semantic output is crucial for informing the subsequent LLM.

\textbf{LLM-assisted Report \& Mitigation Generation:}
In the final stage, the semanticized network descriptions are combined with documented mitigation plans and fed into the LLM via prompt engineering. The LLM, now equipped with an intuitive understanding of the network state, reasons about potential issues and synthesizes comprehensive network reports and tailored mitigation solutions. Furthermore, for specific fault types like configuration errors or attacks, it can generate executable scripts to automate the fault resolution process.
\section{Methodology}
\label{section:methodology}

\subsection{Rule-based Data Scaling }
\label{section:rulebase}
Fig.~\ref{03mutidevice} shows that entity features in HNs exhibit multi-scale properties. To address the challenge of multi-scale KPI features across heterogeneous network entities, we propose a dynamic scaling method based on unsupervised clustering. The core of this approach is to generate a List of Scaled States (LSS) for each entity, which maps raw KPI values with varying scales and dimensions into a unified set of state codes with explicit semantic meanings.\par

The process begins with multi-dimensional feature extraction for each KPI time series. For a given time window T, we compute a feature vector f containing the mean, jitter, variance, and trend, calculated as follows:

\begin{equation}
\label{equation:getkmeans}
\begin{aligned}
f_{avg} &=  \frac{1}{n} \sum_{i=1}^{n} v_i,\\
f_{var} &= \frac{1}{n} \sum_{i=1}^{n} (v_i - f_{avg})^2,\\
f_{jit} &= \sqrt{f_{var}},\\
\end{aligned}
\end{equation}
here, $v$ denotes the discrete data points within the window as $v = {v_1, v_2, ..., v_n}$, and $v_i$ represents the i-th sampled value (e.g., the packet loss rate at time step i).

Trend ($f_{trend}$) is quantified by counting the number of significant local maxima and minima ($N_{extremal}$) within the time window. Specifically, we first subdivide the window $T$ into $m$ overlapping subintervals of equal length. We then identify local extrema by detecting points where the absolute differences with both adjacent values exceed a noise threshold $h$, which is derived from the average variance of historical data under normal conditions. The total count of such significant extrema across all subintervals gives $N_{extremal}$, i.e.,$f_{trend}~=~ N_{extremal}$.

Subsequently, we apply the K-means algorithm to cluster the historical feature vectors $f$ from all entities. The optimal number of clusters $k$ is determined using the elbow method. The clustering yields a set of cluster centers $\{ a_1, a_2, ..., a_k\}$, where each center represents a prototypical network operating mode.

The clustering result inherently reveals the network state distribution. We identify the largest cluster, $A_{normal}$, as the "normal state" baseline, under the assumption that the network spends most of its operational time in a stable regime. The LSS is generated by assigning a state code to each cluster, where the anomaly severity of a state is defined by the Euclidean distance between its cluster center and the baseline center.

The severity $S_j$ for the state corresponding to cluster j is calculated as:

\begin{equation}
\label{equation:SJ}
\begin{aligned}
S_{j} &=  ||a_j - a_{normal}||_2.
\end{aligned}
\end{equation}
\begin{table}[!t]
\caption{Packet Loss Rate Interval Corresponding to Status}
\label{table:statuslist}
\centering
\begin{tabular*}{0.85\columnwidth}{@{\extracolsep{\fill}}c|lcc@{}}
\hline
\diagbox[width=8em]{Entity}{Metrics} & \textbf{Value} & \textbf{Status} & \textbf{Description} \\
\hline
\multirow{2}{*}{City Vehicle} 
& 0\% & 0 & no \\
& 0\%-1\% & 1 & normal \\
& 1\%-5\% & 2 & slight \\
& 5\%-10\% & 3 & moderate \\ 
\hline
\multirow{2}{*}{Expressway Vehicle}
& 0\%-2\% & 1 & normal \\
& 2\%-8\% & 2 & slight \\
& 8\%-15\% & 3 & moderate \\
\hline
\multirow{2}{*}{Plain UAV}
& 0.5\%-1\% & 3 & moderate \\
& 1\%-26\% & 4 & high \\
& 26\%-99\% & 5 & extreme  \\
& 100\% & 6 & complete \\
\hline
\end{tabular*}
\end{table}

\begin{figure}[!t]
\centering
\includegraphics[width=1\columnwidth]{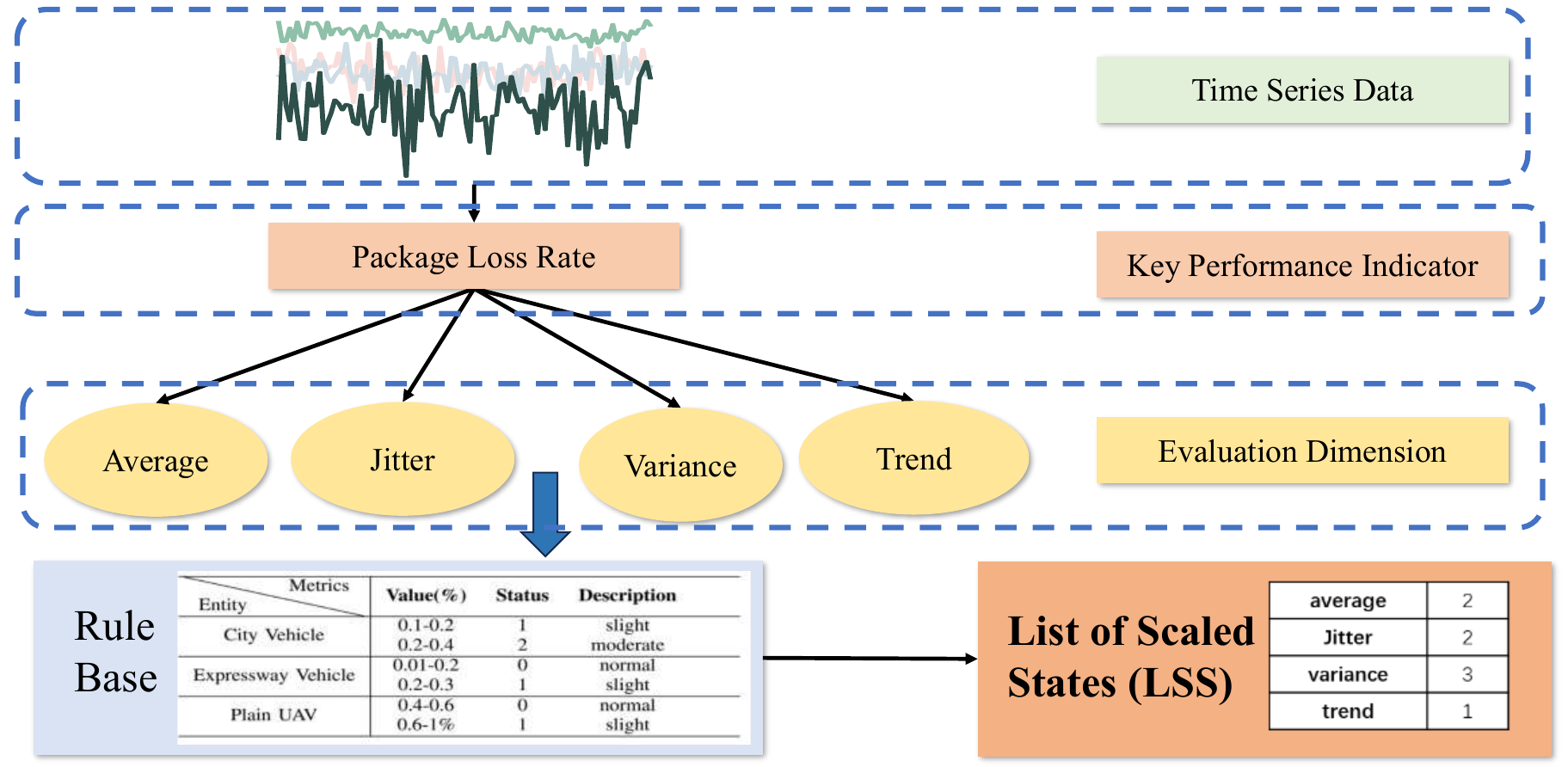}
\caption{ Network node KPI scaling process: MSADM first calculates the mean, variance, jitter, and trend of a set of time series data in the network, and then maps them to corresponding abnormal states through a rule base.}
\label{03datascaling}
\end{figure}

A larger $S_j$ indicates a more severe deviation from normal operation. Finally, for a new KPI time series, its feature vector $f_{new}$ is computed and assigned to the nearest cluster center. The output is the corresponding state code from the LSS. This process dynamically maps a raw KPI value (e.g., a packet loss rate of 0.2\%) to a normalized state (e.g., State Code 1 for "slight anomaly"), based on its comprehensive statistical profile rather than its absolute value alone. \par

Additionally, threshold values across different KPI dimensions hold additional value and significance. Take the packet loss rate as an example. When it falls within the two close numerical intervals of 99\% and 100\%, the fault phenomena presented may be significantly different. The former may imply the existence of malicious traffic, while the latter is more likely to indicate a node crash. Based on this, in our rule base, we not only recorded the intervals obtained through cluster analysis, but also incorporated manually designed intervals. Up to this point, we have constructed a scaling rule base. 

After constructing the rule base,  rule-based maps each cluster defined by its centroid in the feature space to a specific interval of raw KPI values and assigns it a state code. We present some examples of packet loss rate intervals and corresponding state codes in Table~\ref{table:statuslist}.\par

The MSADM scaling process is illustrated in Fig.~\ref{03datascaling}. During its operation, MSADM calculates KPI values based on the detected timing data and derives the LSS from the clustering results that contain these KPI values. Unlike traditional static normalization methods, LSS can dynamically adjust thresholds according to the actual situation of specific entity clusters, thereby achieving consistent scaling among heterogeneous nodes. In the LSS, each state code corresponds to a clear textual description. For example, when the packet loss rate exceeds 5\%, it is assigned to state code 2. When high latency occurs—for instance, reaching 120 milliseconds, it is classified with a higher severity state code, such as 3.\par

\subsection {Network Detection Task}
\label{section:model}

\begin{figure*}[!t]
\centering
\includegraphics[width=0.95\textwidth]{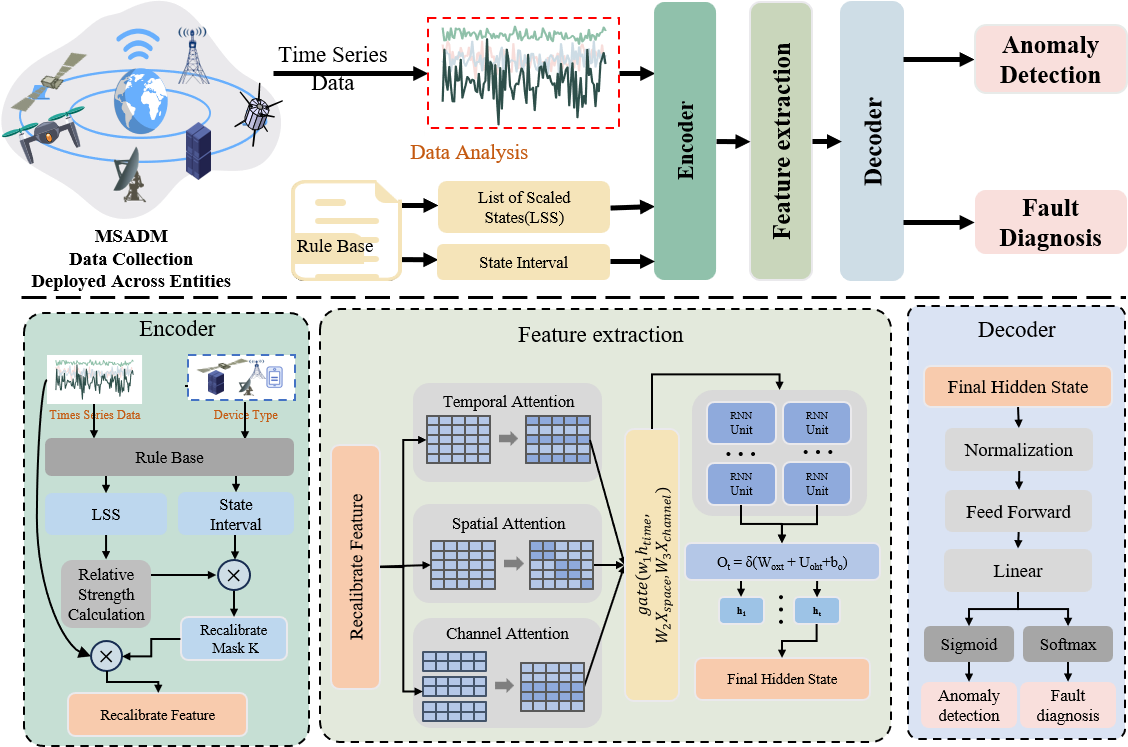}
\caption{ First integrates the anomalous status list with temporal data through weighted fusion. The spatiotemporal attention mechanism subsequently captures dynamic variations in KPI metrics across spatial and temporal dimensions, while the channel attention mechanism discerns feature dependencies among distinct traffic flows. An LSTM layer then models sequential dependencies to learn intrinsic data correlations. Finally, an MLP network jointly performs anomaly detection and fault diagnosis tasks.}
\label{05FaultArchitecture}
\end{figure*}

In this section, we delineate the network detection task, which constitutes the first core function of the MSADM intelligent agent. As illustrated in Fig.~\ref{05FaultArchitecture}, the MSADM detection framework processes the collected time-series data through a structured pipeline comprising three principal components: the LSS-guided feature recalibration encoder, the spatiotemporal-channel attention fusion module, and the Dual-Task decoder.

\textbf{1) Encoder:} While the LSS provides normalized state codes that ensure comparability across heterogeneous entities, it may obscure fine-grained intensity variations within the same anomaly state. For instance, packet loss rates of 30\% and 60\% might both be mapped to the same "severe anomaly" state code, despite their significant operational difference.

To reconcile consistent scaling with the retention of critical intensity information, we design a feature recalibration encoder. This module synthesizes the normalized state code with the relative intensity of the raw KPI value within its assigned state interval. The recalibration process for a feature $f$ is as follows:

First, the state code $s_f$ is retrieved from the LSS, along with its corresponding state interval $I_f = [L_f,U_f]$, where $L_f$ and $U_f$ denote the lower and upper bounds, respectively.

Second, the relative intensity $r_f$ of the current raw value $v_f$ within $I_f$ is calculated:
\begin{equation}
\label{equation:SJ1}
r_f = \frac{v_f - L_f}{U_f - L_f}.
\end{equation}\par

This ratio $r_f \in [0,1]$ indicates the position of the current value within its anomaly severity band, with values closer to 1 representing more extreme manifestations.

Finally, the recalibration weight $K_f$ is computed as the product of the state code and the relative intensity:
\begin{equation}
\label{equation:SJ2}
K_f = s_f \times r_f.
\end{equation}\par
The resulting mask $K = [K_1, K_2,..., K_n]$ is applied to the input features, ensuring the model prioritizes features that are both in a severe anomaly state and at the higher end of their state's intensity spectrum. The recalibrated features $f_v$ are then passed to the subsequent module.

\textbf{2) Feature extraction Module:} The recalibrated features $f_v$ are fed into a hybrid Attention-LSTM architecture designed to capture complex spatio-temporal dependencies. This module incorporates three dedicated attention mechanisms working in concert with an LSTM layer:

Channel Attention: This mechanism adaptively weights different KPIs to identify feature-level importance, allowing the model to focus on critical metrics such as sudden spikes in packet loss across multiple connections.

Spatial Attention: This mechanism models dynamic interactions between network entities while detecting abrupt temporal variations, capturing correlated anomalies indicative of cascading failures.

Temporal Attention: This mechanism emphasizes important time steps within the sequence, highlighting periods of significant performance degradation or recovery.

Rather than simply concatenating these attention outputs, we propose a \textit{Gated Fusion Module} to adaptively integrate the three representations $X_{chan}, X_{spat},  X_
{temp}$. The fusion process begins by projecting each representation to a common feature space and summing them:

\begin{equation}
\mathbf{X}_{\text{sum}} = \mathbf{W}_1 \times \mathbf{X}_{\text{temp}} + \mathbf{W}_2 \times \mathbf{X}_{\text{spat}} + \mathbf{W}_3 \times \mathbf{X}_{\text{chan}},
\end{equation}
where, $W_1,~W_2,~and~W_3$ are all trainable parameters.

A gating signal is then generated to control the information flow:
\begin{equation}
\mathbf{g} = \sigma(\mathbf{W}_g \mathbf{X}_{\text{sum}} + \mathbf{b}_g).
\end{equation}

The final fused representation is obtained through gated weighting:
\begin{equation}
\mathbf{X}_{\text{fused}} = \mathbf{g} \odot \mathbf{X}_{\text{sum}}.
\end{equation}

This gated fusion enables the model to dynamically emphasize the most salient features—prioritizing spatial correlations during cascading failures while focusing on temporal patterns during gradual performance degradation.

The fused representation $\mathbf{X}_{\text{fused}}$ is then fed into the LSTM to model temporal dependencies, with its final hidden state $\mathbf{h}_T$ serving directly as the comprehensive feature representation $\mathbf{attr}$ for subsequent tasks.

\textbf{3) Decoder:} The decoder integrates the spatio-temporal representation $attr$ to form a high-level anomaly representation 
$I_sa$:
\begin{equation}
\label{sa}
\mathbf{I}_{sa} = f_1(\mathbf{W} \times \mathbf{attr} + \mathbf{b}),
\end{equation}
where $W,~b$ are trainable parameters, and 
$f_1$ is the ReLU activation function.

Finally, $I_{sa}$is fed into two parallel fully connected layers to perform anomaly detection and fault classification simultaneously. The total loss L is a weighted sum of the detection loss $L_d$ and the classification loss $L_c$:
\begin{equation}
\label{equation:loss}
\begin{aligned}
\mathcal{L}_d &= -\sum{i} y_{di} \log(p_{di}),\\
\mathcal{L}_c &= -\sum{i} y_{ci} \log(p_{ci}),\\
\mathcal{L} &= \kappa \mathcal{L}_d + (1 - \kappa) \mathcal{L}_c,
\end{aligned}
\end{equation}
where $y_d, y_c$ are the true labels for anomaly detection and fault diagnosis, $p_d, p_c$ are the predicted probabilities, and $k$ is a task-weighting hyperparameter. The model parameters are optimized by minimizing $L$.

\subsection{Network Information Semanticization}
\label{section:semantictree}
\begin{figure}[!t]
\centering
\includegraphics[width=0.9\columnwidth]{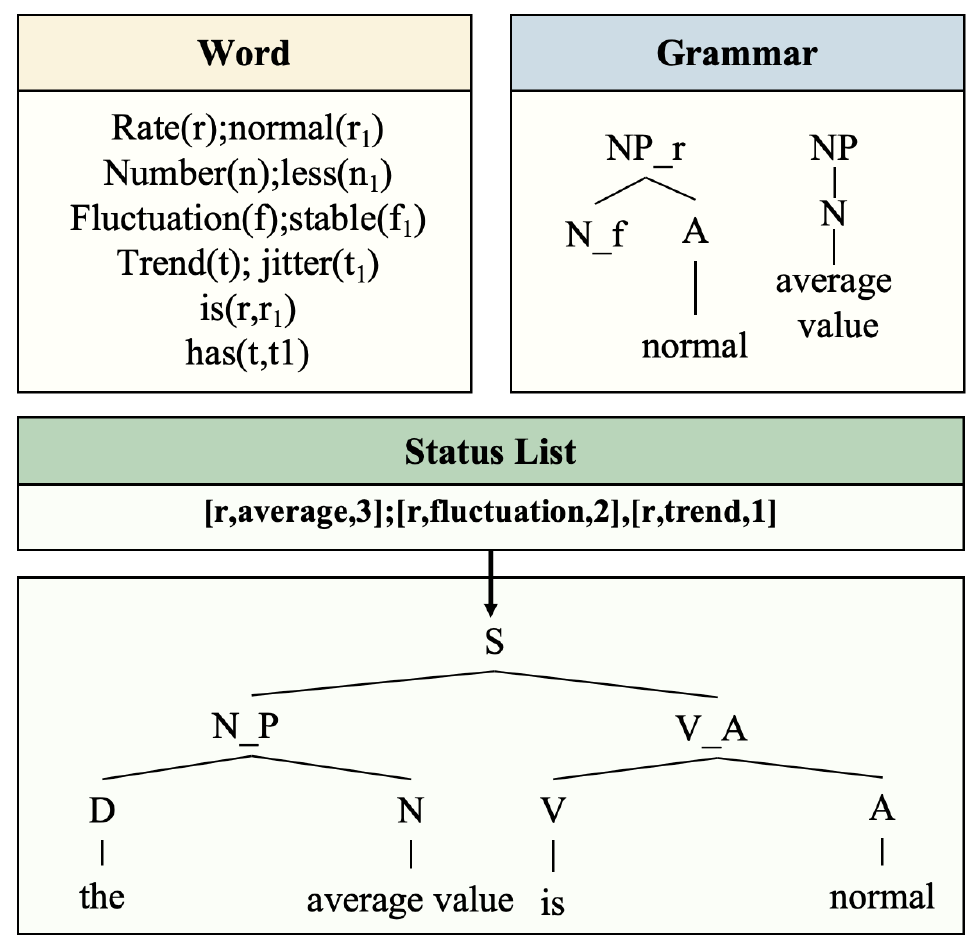}
\caption{Semantic description generation framework. The semantic rule tree, built on a context-free grammar, translates state codes into natural-language descriptions via deterministic path traversals.}
\label{06ExceptionReportGeneration}
\end{figure}

This module serves as a semantic bridge between the quantitative anomaly detection results and the qualitative reasoning capability of LLMs. It systematically converts the numerical state codes from Section~\ref{section:rulebase} into structured natural language descriptions, providing interpretable context for downstream mitigation tasks.

We formalise the description generation process using a \emph{Semantic Rule Tree}, defined as a rooted directed tree $T=(V, E)$, where $V$ represents vertices (words/phrases) and $E$ denotes grammatical relations. The tree is constructed with a context-free grammar backbone. Each complete sentence corresponds to a path $P$ from the root $r$ to a leaf node $l$, such that $P = (r, v_1, v_2, ..., l)$.

The vocabulary for leaf nodes, particularly the state descriptors $D$, is populated through a direct mapping function $M$ from the scaled state codes $S_c$:
\begin{equation}
\label{equation:MSC}
M: ~S_c \rightarrow D.
\end{equation}\par
For instance, for packet loss rate KPI: $M(1)=\text{``normal''}$, $M(2)=\text{``slight''}$, $M(3)=\text{``moderate''}$.

Given an input vector of state codes $\mathbf{S} = [S_{c1}, S_{c2}, ..., S_{cn}]$ for $n$ KPIs, the generation follows a deterministic procedure. The algorithm utilises two core functions: $\text{TraverseTree}(T, d_i)$ identifies the unique path to the leaf node containing descriptor $d_i$, while $\text{ConstructSentence}(path_i)$ concatenates nodes along this path into a grammatically correct sentence.

The semanticization process inherits adaptability from the underlying rule base. Periodic updates to the rule base through:
\begin{equation}
\label{equation:Recluster}
\theta^{(t+1)} = \text{Recluster}(\mathcal{D}_{\text{recent}}^{(t)}),
\end{equation}
where $\mathcal{D}_{\text{recent}}^{(t)}$ represents recent network data, automatically induce updated mapping $M^{(t+1)}$. This ensures semantic labels remain consistent with evolving network patterns without modifying the tree structure $T$.

To enhance conciseness, we implement two optimisation strategies.

Normal State Pruning: all descriptions where $S_{ci} = S_{\text{normal}}$ are filtered, significantly reducing input length to the LLM while maintaining diagnostic relevance.

Conditional Value Reporting: for severe anomalies where the raw value $v$ exceeds the state interval upper bound ($v > \text{upper} \times \tau$, with $\tau=1.15$), the specific value is appended to provide critical diagnostic details.

\begin{algorithm}[!t]
  \caption{Enhanced Semantic Description Generation}
  \label{alg:semantic}
  \renewcommand{\algorithmicrequire}{\textbf{Input:}}
  \renewcommand{\algorithmicensure}{\textbf{Output:}}
  \begin{algorithmic}[1]
    \Require State code vector $\mathbf{S}$, Semantic rule tree $T$, Mapping $M$, Recent data $\mathcal{D}_{\text{recent}}$, Time period $\Delta t$
    \Ensure Sentence list $L$
    \State $L \leftarrow [~]$ 
    \State $update\_flag \gets \text{CheckUpdateTime}(\Delta t)$
    \If {$update\_flag$}
        \State $\theta \gets \text{Recluster}(\mathcal{D}_{\text{recent}})$
        \State $M \gets \text{UpdateMapping}(\theta)$
    \EndIf
    \For{each state code $S_{c_i} \in \mathbf{S}$}
      \If {$S_{c_i} \neq S_{\text{normal}}$} 
        \State $d_i \gets M(S_{c_i})$
        \State $path_i \gets \text{TraverseTree}(T, d_i)$ 
        \State $sentence_i \gets \text{ConstructSentence}(path_i)$
        \State $I \gets \text{GetStateInterval}(S_{c_i})$
        \If {$v_i > I.upper \times 1.15$}
            \State $sentence_i \gets sentence_i + \text{`` with value } v_i \text{''}$
        \EndIf
        \State $L.\text{append}(sentence_i)$
      \EndIf
    \EndFor
    \State \Return $L$
  \end{algorithmic}
\end{algorithm}

Algorithm~\ref{alg:semantic} details the enhanced semantic description generation process. The algorithm first checks if a rule base update is needed based on the time period $\Delta t$ (lines 2-6). For each non-normal state code, it retrieves the corresponding semantic descriptor (line 8), generates the sentence through tree traversal (lines 9-10), and applies conditional value reporting for severe anomalies (lines 11-13). This integrated approach ensures the generated descriptions are both concise and informative while maintaining adaptability to network dynamics.

\subsection{LLM Assisted Solution Generation}
\label{section:llm} 
We input the anomaly detection results obtained from MSADM, along with the generated text prompts, into the LLM to obtain a network report and anomaly solution. LLM's powerful natural language processing capabilities enable it to deeply understand semantic information and derive meaningful features and patterns~\cite{10591707}. Moreover, LLMs' continuous learning ability enables them to adapt and effectively respond to changing event types, showing excellent scalability and rapid adaptability in complex scenarios. However, LLMs themselves suffer from limitations in reasoning, hallucination, and input length. We need to process the anomalous information before inputting it into the LLM to solve these problems. \par
Regarding the issue of hallucinations in reasoning, since LLMs are based on patterns in data rather than genuine understanding, they often encounter difficulties when performing complex and in-depth reasoning. This makes it challenging for them to consistently generate accurate and contextually consistent responses, which require profound domain knowledge or logical coherence. During the integration process, we combine the anomaly detection results and semantic results from MSADM to guide the LLM, with the aim of assisting in generating anomaly reports that better meet the requirements of network management.

\begin{figure}[!t]
\centering
\includegraphics[width=0.9\columnwidth]{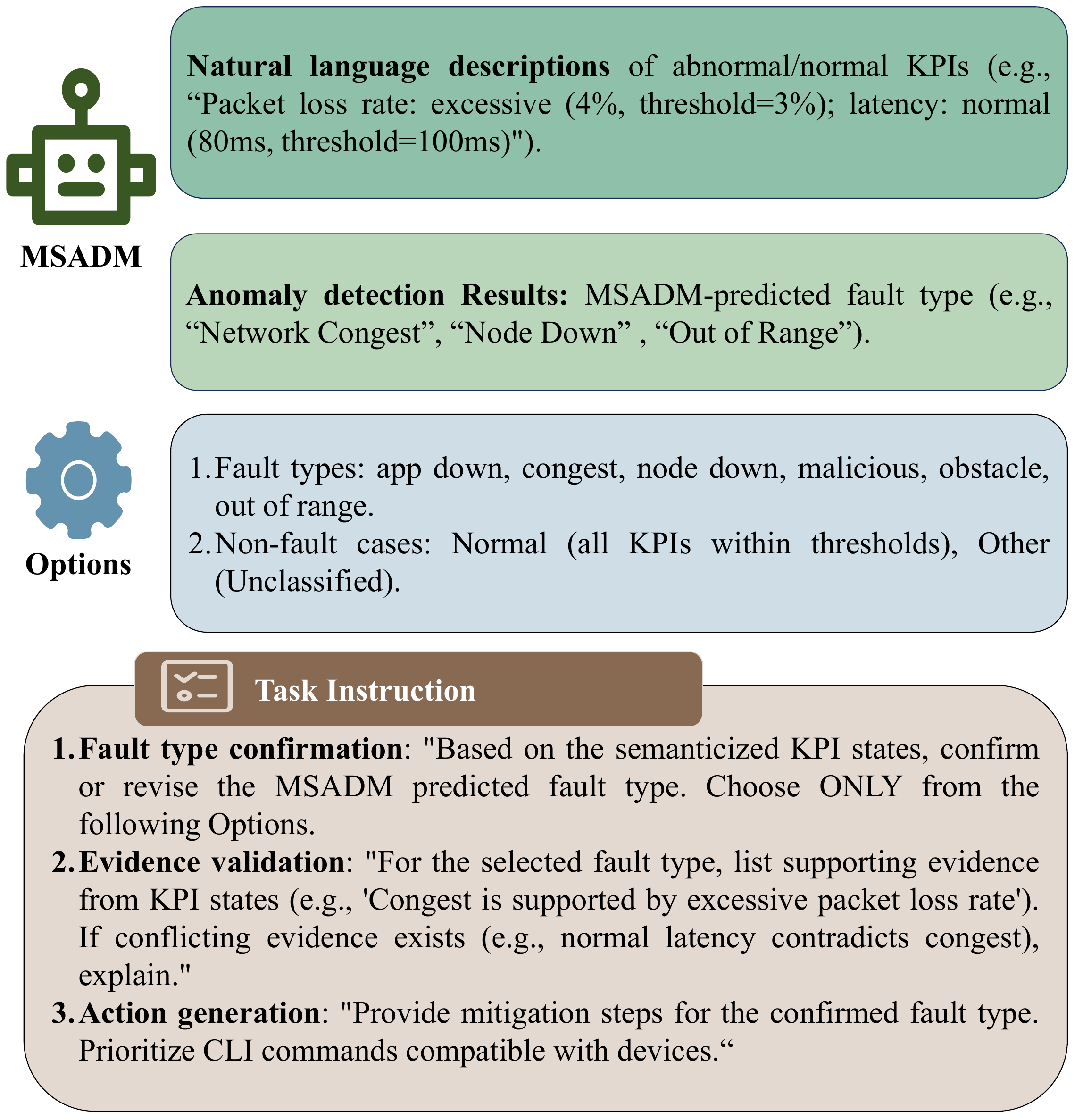}
\caption{Partial prompt template display: The prompt architecture combines MSADM's anomaly severity descriptions with its detection results, adds common anomalies to input to reduce LLM-induced output anomalies, and enforces structured tasks: verify diagnostics via factual checks, explain root causes through causal reasoning, and propose actionable fixes within network constraints. }
\label{07texttip}
\end{figure}

\begin{figure}[!t]
\centering
\includegraphics[width=0.95\columnwidth]{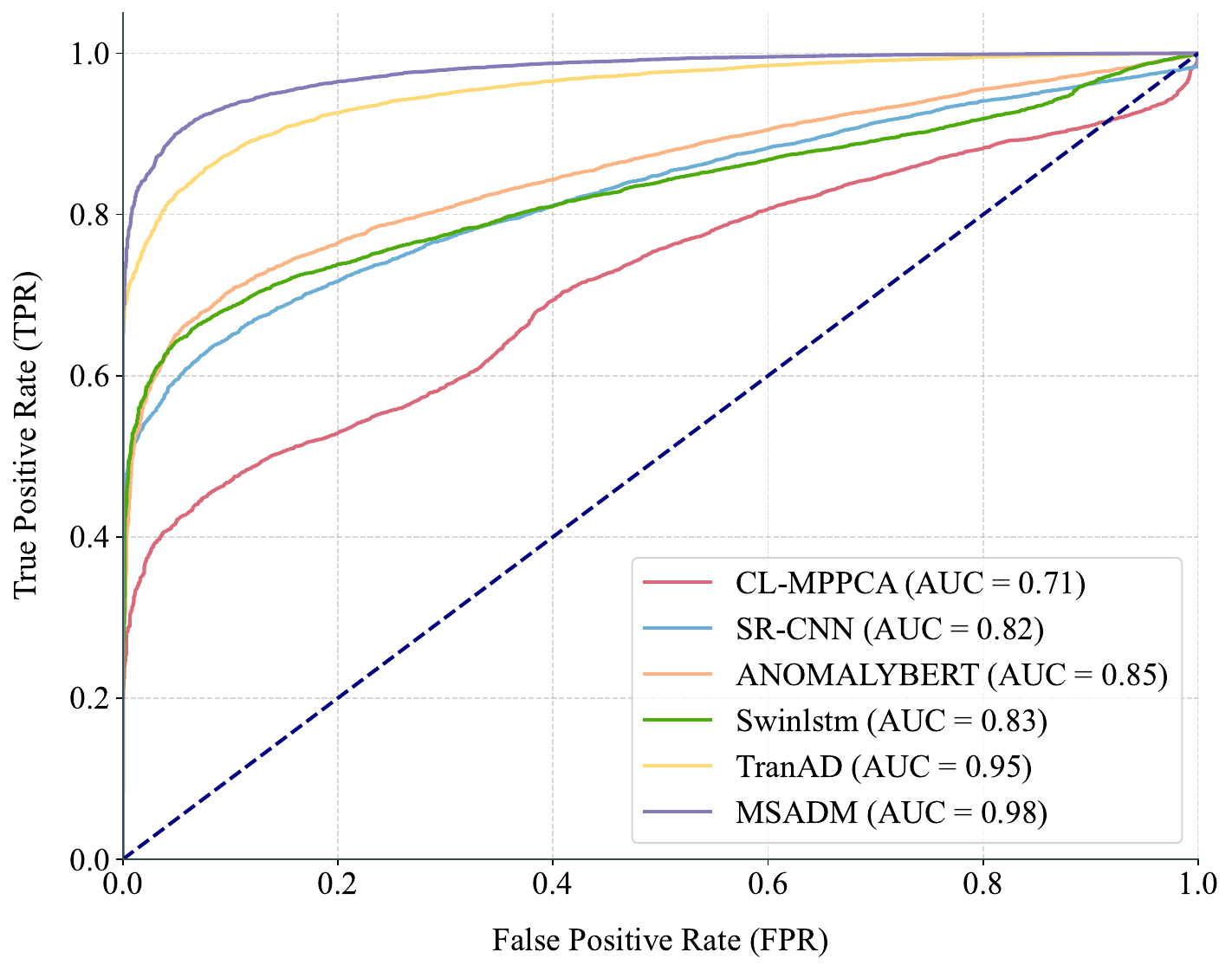}
\caption{Comparative Analysis of ROC Curves for Multi-Model Outlier Detection.}
\label{roc}
\end{figure}

\textbf{Prompt Structure:} Our prompt structure (Fig.~\ref{07texttip}) enforces strict schema alignment between network context and mitigation options. Unlike generic prompt designs, the prompt includes three fixed segments: \par

\textbf{Context}: It includes the fault prediction results from MSADM, along with the textual description of the KPI status of the current node.\par
\textbf{Options}: We provide restricted output options to prevent speculative conclusions. These options include the "normal" and "unclassified" categories for edge cases.\par
\textbf{Task Instruction}: The tasks for the LLM are defined in three aspects: fault type confirmation, evidence verification, and action generation. \par  

The LLM delivers two outputs for heterogeneous network health management. At first, it generates a network anomaly report detailing identified fault types, their severity levels, and supporting semanticized KPI evidence. This evidence explicitly highlights patterns within the KPIs, for example, indicating conflicting or supporting trends that diagnose the issue. The second, the system provides actionable mitigation plans. For rule-based resolvable issues, this includes automated API-driven corrective actions, such as limited traffic. For complex faults requiring human expertise, it furnishes administrator-oriented guidance; for instance, providing specific CLI commands or detailing vendor escalation protocols. \par
To ensure reliability, our structured prompting framework enforces a Chain-of-Verification methodology~\cite{Fu2022ComplexityBasedPF}, which decomposes the reasoning process into sequential steps – fault type confirmation from constrained options, evidence validation against semantic states, and context-aware action generation. \par

\begin{table*}[!t]
\caption{Performance comparison of different models with data-scaling}
\label{perforcom}
\centering
\begin{tabular*}{\textwidth}{@{\extracolsep{\fill}}l *{4}{c} *{4}{c} } 
\hline
\multirow{2}{*}{\diagbox{Model}{Metrics}} &\multicolumn{4}{c}{Anomaly Detection} & \multicolumn{4}{c}{Fault Diagnosis} \\ 
\cline{2-5} \cline{6-9}
& Accuracy$ \uparrow$ & Recall$ \uparrow$ & FNR$ \downarrow$ & FPR$ \downarrow $&Accuracy$ \uparrow$ &Recall$ \uparrow$ &FNR$ \downarrow$ &FPR$ \downarrow$ \\
\hline
{SR-CNN}&85.46&90.66&9.34&46.79&63.25&71.67 &28.33 &39.28\\
{CL-MPPCA}&87.19&91.22&8.78&41.96&67.61&73.24&26.76&37.41\\
{ANOMALYBERT}&91.39&96.82&3.18&42.47&68.38& 71.46 & 28.54 & 37.23  \\
{Swinlstm}&87.59&89.16&10.84&22.23&78.24&81.47&18.53&24.34\\
{TranAD}&89.1&91.51&8.49&25.95&81.15&84.26&15.74&21.88\\
\hline
{MSADM}&\textbf{97.09}&\textbf{98.14}&\textbf{1.86}&\textbf{15.52}&\textbf{89.42}&\textbf{91.79}&\textbf{8.21}&\textbf{16.91}\\
\hline
\end{tabular*}
\end{table*}


\section{Evaluation}
\label{section:shiyan}
In this section, we evaluate MSADM through comprehensive experiments addressing four critical aspects.

\textbf{Q1:} Can MSADM surpass other methods in both fault diagnosis and security attack detection? \par 

\textbf{Q2:} Is every component of MSADM important?\par

\textbf{Q3:} Compared to extraction and abstraction methods, can semantic rule trees produce higher-quality text? \par

\textbf{Q4:} How effectively can LLM-generated reports guide network maintenance actions to restore QoS targets? \par

\subsection{Experiment Settings}
\textbf{Dataset Description:} We experiment on two datasets: \par
1) We employ the dataset~\cite{10806770} to conduct the entire process of network health management. This dataset collects communication network data under seven distinct scenarios. It comprises 40 features and has accumulated nearly 20,000 data entries, all of which have been meticulously annotated. Finally, we have released an open-source demonstration of MSADM to elucidate this workflow \footnote{https://github.com/SmallFlame/MSADM}. \par
2) NF-CSE-CIC-IDS2018~\cite{10515433} is a widely-used benchmark dataset for evaluating intrusion detection systems. This dataset simulates a real-world network environment by executing various attacks within a hybrid architecture that integrates modern enterprise applications and network protocols. Its traffic includes normal behavior as well as seven major attack types, specifically: Distributed Denial of Service (DDoS) attacks, Denial of Service (DoS) attacks, brute force attacks, botnet activities, infiltration attacks, and web-based attacks (such as SQL injection and cross-site scripting). The dataset contains over 16 million records in total, with attack traffic accounting for approximately 17\%.\par
\textbf{Baselines: }We divide baselines into three categories. One is an anomaly detection model, one is an attack detection model, and one is a text generation model. \par
1) Anomaly Detection Model: It's mostly based on machine learning anomaly detection models, including SR-CNN~\cite{ren2019time} and CL-MPPCA~\cite{tariq2019detecting}, and 
AnomalyBERT~\cite{jeong2023anomalybert} and Swinlstm~\cite{tang2023swinlstm} and TranAD~\cite{tuli2022tranad}.\par
2) Attack Detection Model: our experiment includes widely used XGBoost~\cite{technologies13030088}, AlexNet~\cite{zhao2025research}, miniVGGNet~\cite{serdaly2024improving}.\par
3) Text Generation Model:
We compare extraction-based methods such as CTR~\cite{wang2011collaborative}, NARRE~\cite{chen2019co} a nd abstraction-generation-based methods such as NRT~\cite{10.1145/3077136.3080822}, KEGNN~\cite{lyu2022knowledge}. \par
\textbf{Detection Metric:} To evaluate the anomaly detection and attack detection performance of MSADM and baselines, our evaluation involved several widely used metrics: Accuracy, Recall, False Negative Rate, and False Positive Rate. \par
\textbf{Text Quality Metric:} The quality of the generated text is assessed using both automatic metrics and expert evaluation. ROUGE measures the n‑gram overlap between the generated text and the reference, reflecting content coverage. BLEU evaluates the precision of n‑gram matches, focusing on the fluency and formulation accuracy. In addition, an Expert Score is provided by domain specialists to assess the technical correctness and practical utility of the generated reports. All metrics are scaled from 0 to 100, with higher scores representing better performance.



\begin{figure*}[!t]
\centering
\subfloat[ Anomaly Detection Accuracy]{\includegraphics[width=0.9\columnwidth]{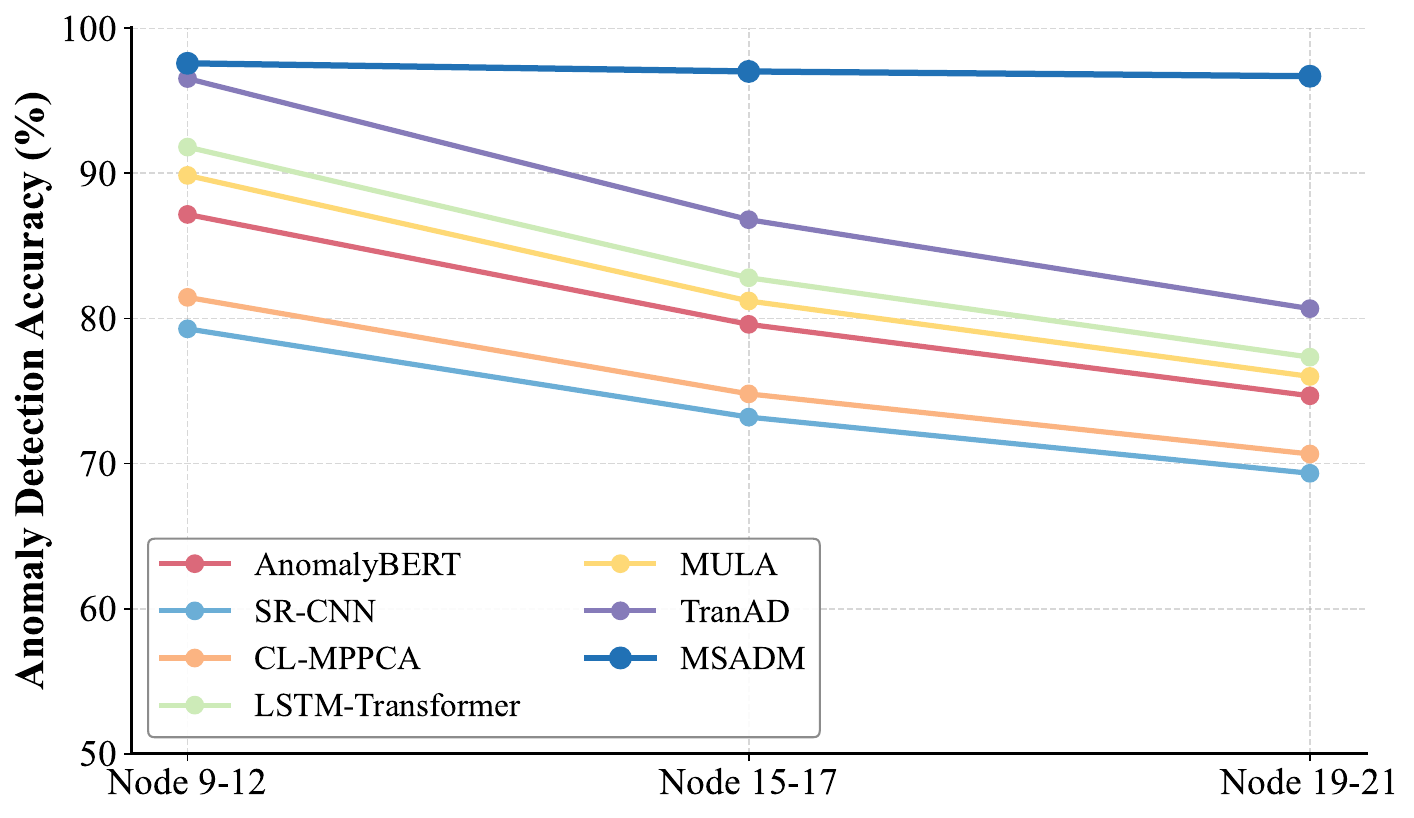}}
\hfill
\subfloat[ Fault Diagnosis Accuracy]{\includegraphics[width=0.9\columnwidth]{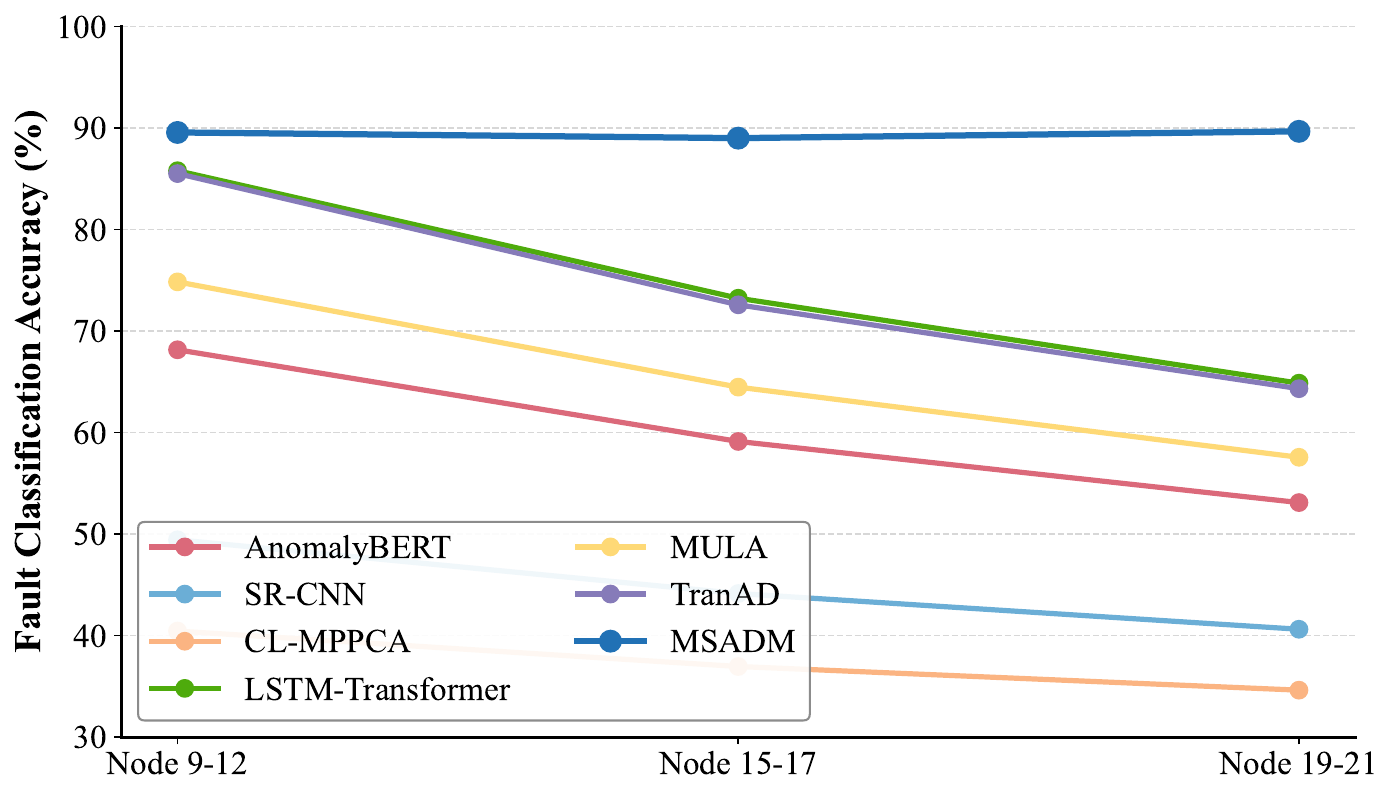}}
\caption{The comparison is performed under the network with the number of nodes being 9-12, the number of nodes being 15-17, and the number of nodes being 19-21. }
\label{multi-node-acc}
\end{figure*}

\subsection{MSADM vs SOTA in Fault and Attack Detection}

To answer Q1, we conducted experiments and evaluated the capabilities of MSADM from the following perspectives.\par
\textbf{The importance of data scaling was evaluated by accuracy and recall rate:} In Table~\ref{perforcom} with multi-scale normalisation, MSADM outperforms SOTA models in most metrics, achieving 97.09\% accuracy and 1.86\% FNR in anomaly detection, and 89.42\% accuracy in fault diagnosis. \par

\textbf{ROC comparison:} The ROC curves represent the true positive rate (TPR) and false positive rate (FPR) under different threshold settings. To compare the robustness and reliability of the models. We plotted the ROC curve. As shown in Fig.~\ref{roc}, the ROC curve of the MSADM model is higher than that of other models most of the time. \par




\textbf{Multi-node Architecture:} Due to the anomaly's limited range of influence, enlarging the network size might result in overlooking the anomaly. Fig.~\ref{multi-node-acc} illustrates the variation in model accuracy corresponding to changes in network size. In both scenarios with a small and large number of nodes, the MSADM model outperforms other models in both anomaly detection and classification accuracy. \par



\begin{table}[!t]
\caption{ Detection Accuracy OF MSADM and Baselines on NF-CSE-CIC-IDS2018}
\label{NF-CSE-CIC-IDS2018}
\centering
\begin{tabular}{@{\extracolsep{\fill}}l *{2}{c} *{2}{c} } 
\hline
\multirow{2}{*}{\diagbox{Model}{Metrics}} &\multicolumn{2}{c}{Anomaly Detection} & \multicolumn{2}{c}{Attack Detection} \\ 
\cline{2-3} \cline{4-5}
& Accuracy$\uparrow$ & Recall$ \uparrow$ & Accuracy$ \uparrow$ &Recall$ \uparrow$  \\
\hline
{XGBoost}&96.03&97.99& 93.81 & 94.49\\
{AlexNet}&98.4&98.13&93.24& 94.23  \\
{miniVGGNet}&98.7&96.99&93.66&93.66\\
\hline
{MSADM}&\textbf{99.34}&\textbf{99.65}&\textbf{94.27}&\textbf{95.68}\\
\hline
\end{tabular}
\end{table}

\begin{table*}[!t]  
\caption{Performance comparison of different models} 
\label{table:gpu}  
\centering  
\begin{tabular*}{\textwidth}{@{\extracolsep{\fill}}l *{1}{c} *{3}{c} *{3}{c}} 
\hline  
\multirow{2}{*}{Method} & \multicolumn{1}{c}{FLOPs (M)} & \multicolumn{3}{c}{GPU Utilization (\%)} & \multicolumn{3}{c}{Inference Time (ms)} \\  
\cline{2-2} \cline{3-5} \cline{6-8}  
& Value 
& Best & Average & Variance
& Best & Average & Variance\\  
\hline  
SR-CNN   & 8.62 & 11.14 & 14.53 & 2.1 & 7.23 & 9.15 &  2.57  \\  
CL-MPPCA & 9.49 & 13.79 & 14.55 & 1.86 & 8.45 & 10.27 & 3.86 \\  
ANOMALYBERT  & 13.26 & 14.91 & 16.8 & 2.61 & 8.97 & 17.62 & 25.15  \\    
Swinlstm &248.9 & 21.82 & 30.64 & 8.4 & 23.48 & 39 & 19.45 \\   
TranAD & 548.7 & 30.82 & 48.64 & 20.61 & 28.96 & 49.71 & 30.17  \\   
MSADM & 36.58  & 16.15 & 18.64 & 2.69 & 10.17 & 19.41 & 14.61 \\
\hline  
\end{tabular*}  
\end{table*}

\textbf{Comparative Analysis on  NF-CSE-CIC-IDS2018:} Given that the dataset contains a significantly smaller volume of attack-type data compared to the abundant normal data, we performed downsampling on the normal dataset. Meanwhile, for Web Attack, which has the least amount of data, we employed SMOTE's oversampling technique to augment the data volume, ensuring that each attack type has an equivalent amount of data. To further validate the generalizability of our model and enhance the credibility of experimental conclusions, we conducted supplementary experiments on widely recognized network attack datasets. Due to the lack of sufficient network attack-type samples, we applied the SMOTE technique to balance the samples used.  \par
As shown in Table~\ref{NF-CSE-CIC-IDS2018}, we compared the performance of the MSADM model in attack detection scenarios, and it outperformed existing methods across all evaluation metrics.

We compare the computational efficiency of the proposed MSADM model against baseline methods, as summarized in Table \ref{table:gpu}. In terms of computational complexity, to ensure a fair comparison focused on anomaly detection performance, all reported FLOPs are evaluated under the lightweight local model setting, excluding the subsequent LLM-based generation stage. Under this setting, MSADM requires only 36.58 M FLOPs, which is significantly lower than Swinlstm and TranAD. Regarding GPU utilization, MSADM achieves an average of 18.64\% with a low variance of 2.69\%, indicating stable and moderate resource usage compared to most baselines. For inference time, MSADM maintains an average of 19.41 ms, outperforming ANOMALYBERT, Swinlstm, and TranAD, while its low variance demonstrates consistent inference speed. Overall, MSADM achieves favorable computational efficiency with lower local complexity, stable resource utilization, and faster, more consistent inference.

\subsection{Ablation experiment}

\begin{figure*}[!t]
\centering
\subfloat[ Ablation Study on Classification ]{\includegraphics[height=6cm,width=0.48\linewidth]{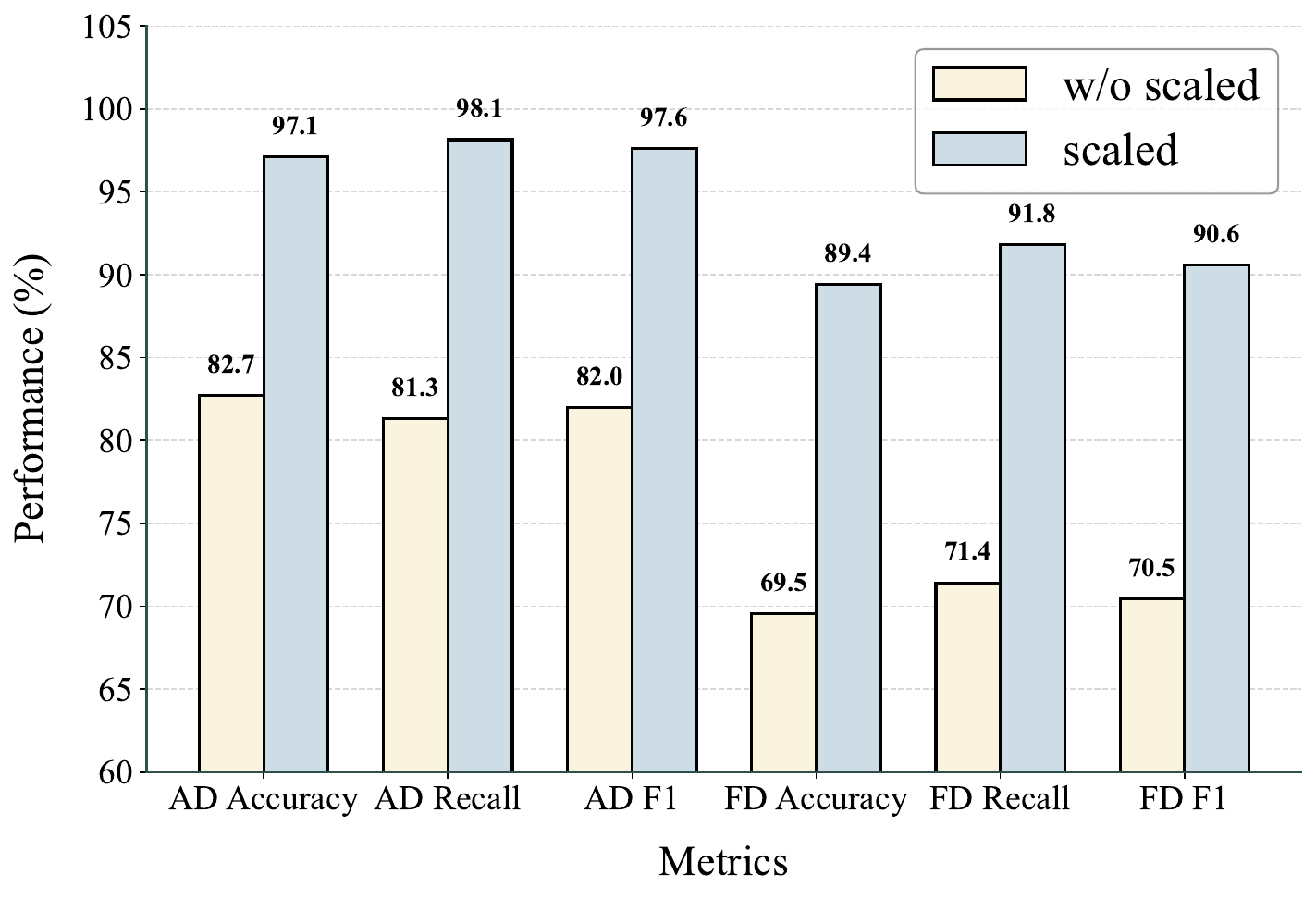}}
\hfill
\subfloat[Ablation Study on Text Generation]{\includegraphics[height=6cm,width=0.48\linewidth]{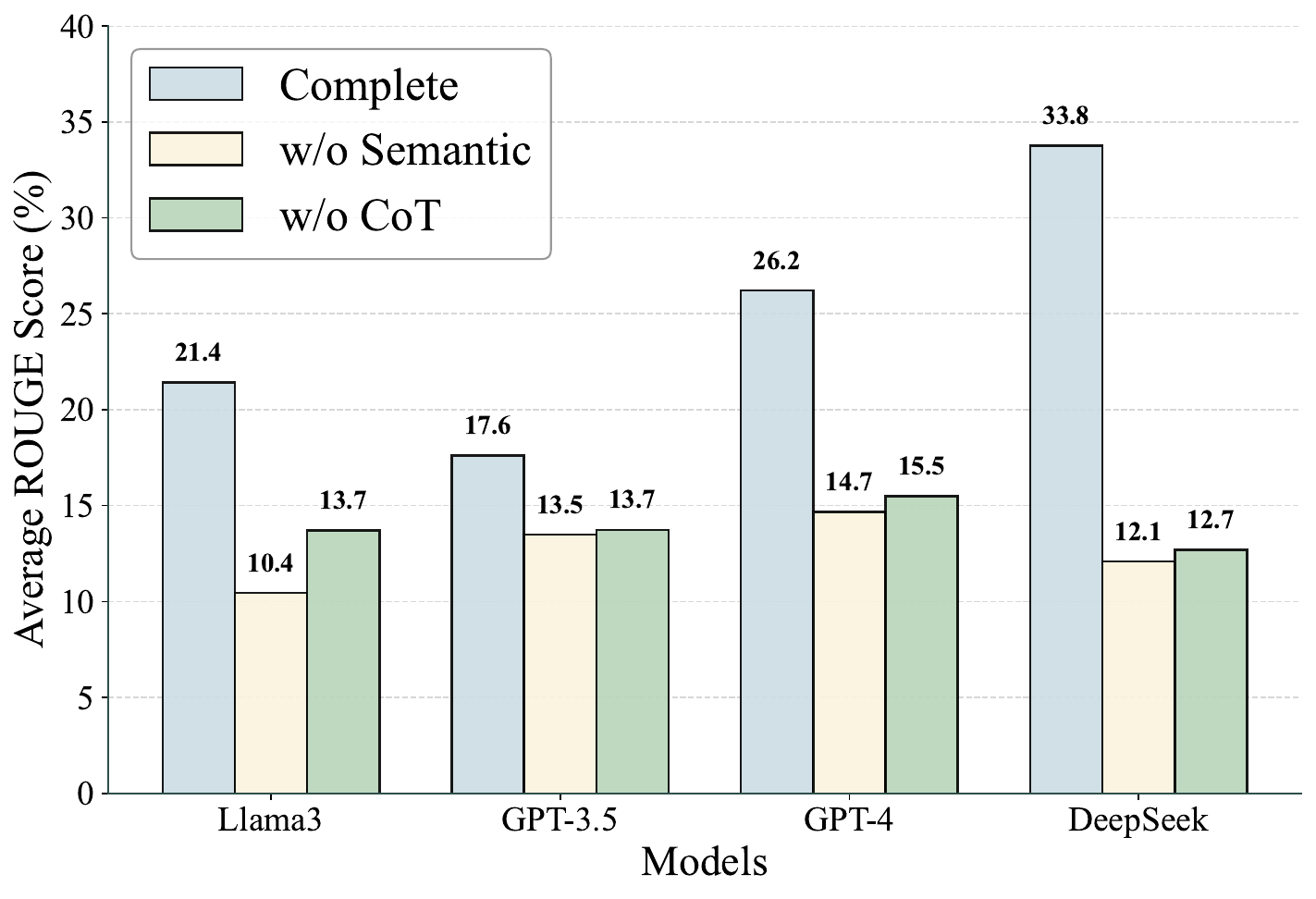}}
\caption{The ablation experiments of standardized and semantic components are presented. We presented the performance of MSADM in terms of accuracy, recall rate, and F1 score both without and after standardization. In addition, we conducted semantic ablation experimental tests on multiple LLMS. }
\label{llm-perforcom-noyuyi}
\end{figure*}
To address research Q2 and validate the effectiveness of each component in the proposed method, we conducted ablation studies on three key modules: data scaling, semanticization, and chain-of-thought (CoT) prompting. The experimental results are shown in Fig.~\ref{llm-perforcom-noyuyi}.

In the ablation study on data scaling, we compared the impact of applying scaling versus not applying scaling (w/o scaled) on the accuracy of anomaly detection and fault diagnosis. The results show that data scaling significantly improves model performance. The average detection accuracy increased from 69.5\% without scaling to 91.8\% with scaling, an improvement of over 22\%. This confirms that data scaling effectively mitigates inconsistencies and dimensional variations in the raw data, providing a unified input distribution that enhances the model's detection and diagnostic capability across different communication nodes.

In the ablation study on semanticization and CoT (Fig.~\ref{llm-perforcom-noyuyi}), we compared the average ROUGE scores across multiple LLM under three settings: the complete model, without semanticization (w/o Semantic), and without CoT (w/o CoT). The results demonstrate that the semanticization module contributes most significantly to performance improvement. Particularly for the DeepSeek model, the ROUGE score of the complete model reached 26.2\%, while removing semanticization caused a drop to 14.7\%. The CoT module also plays an important role in output quality. For Llama3 and GPT-4 models, removing CoT led to a decrease in scores from 21.4\% to 13.7\% and from 17.6\% to 13.5\%, respectively. This indicates that semantic input provides more interpretable and task-consistent representations, while CoT enhances textual coherence and structural integrity through step-by-step reasoning.

\begin{table*}[!t]  
\caption{ROUGE Evaluation and Inference Time of Different Text Generation Models} 
\label{gentext-perforcom}  
\centering  
\begin{tabular*}{\textwidth}{@{\extracolsep{\fill}}l *{3}{c} *{3}{c} *{3}{c} c} 
\hline  
\multirow{2}{*}{Method} & \multicolumn{3}{c}{ROUGE-1} & \multicolumn{3}{c}{ROUGE-2} & \multicolumn{3}{c}{ROUGE-L} & \multirow{2}{*}{Inference Time (ms) $\downarrow$} \\  
\cline{2-4} \cline{5-7} \cline{8-10}  
& Recall$ \uparrow$ & Precision$ \uparrow$& F1$ \uparrow$ & Recall$ \uparrow$ & Precision$ \uparrow$ & F1$ \uparrow$ & Recall$ \uparrow$ & Precision$ \uparrow$ & F1$ \uparrow$ &  \\  
\hline  
CTR & 14.93 & 11.14 & 12.76 & 2.36 & 1.96 & 2.14 & 12.74 & 10.45 & 11.48 & 45.2 \\   
NARRE & 12.54 & 9.23 & 10.63 & 1.86 & 1.64 & 1.74 & 10.84 & 8.95 & 9.80 & 38.7 \\   
NRT & 16.26 & 17.25 & 14.99 & 2.61 & 3.07 & 2.82 & 12.27 & 14.96 & 13.48 & 52.1 \\   
KEGNN & 16.94 & 17.82 & 15.64 & 2.57 & 3.27 & 2.88 & 12.68 & 15.83 & 14.08 & 61.5 \\   
MSADM & \textbf{27.19} & \textbf{51.67} & \textbf{35.63} & \textbf{10.30} & \textbf{21.25} & \textbf{13.87} & \textbf{26.32} & \textbf{34.09} & \textbf{29.71} & \textbf{32.3} \\
\hline  
\end{tabular*}  
\end{table*}

\begin{figure}[!t]
\centering
\includegraphics[width=1\columnwidth]{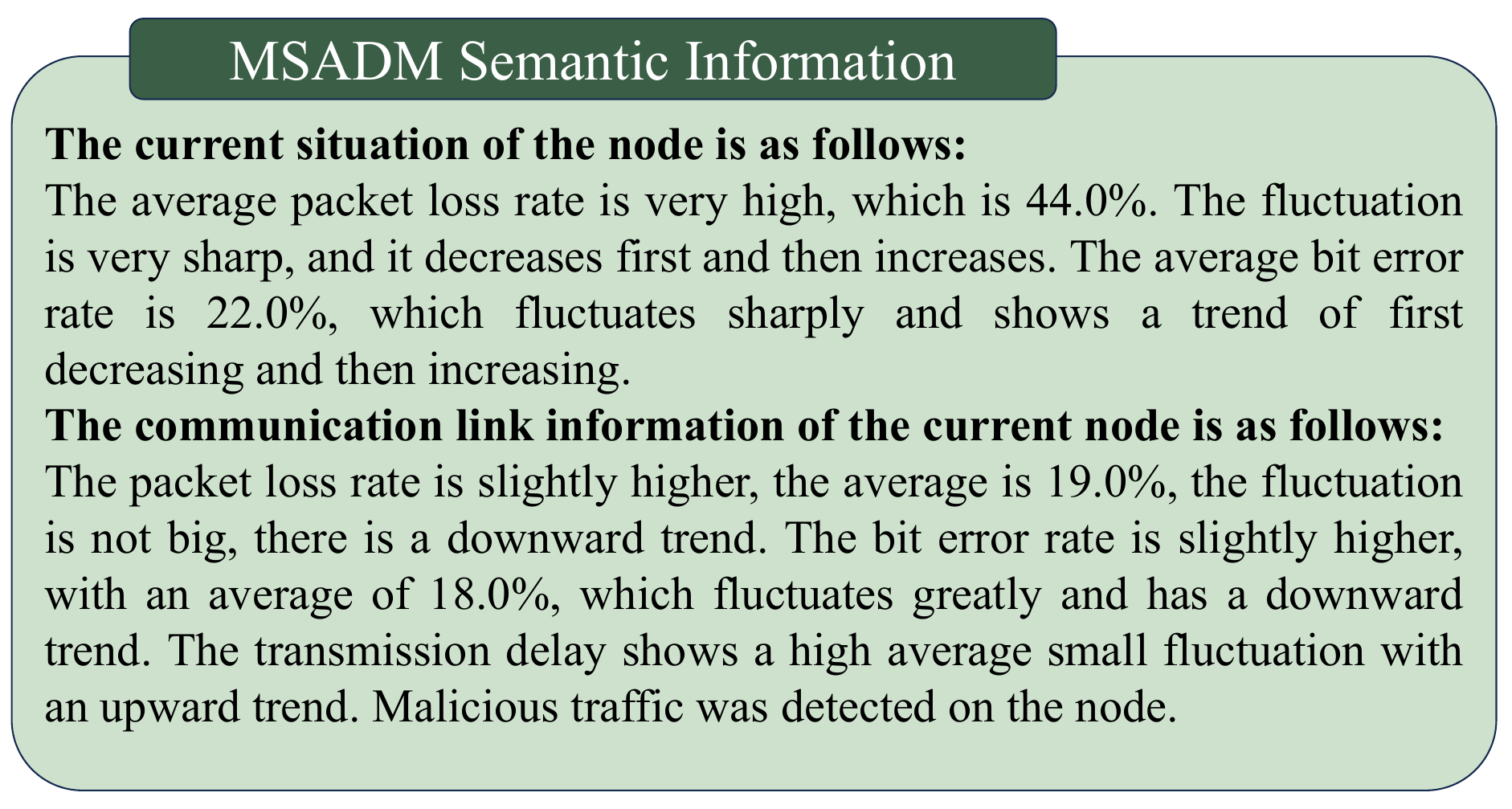}
\caption{MSADM semantic Information. }
\label{msadm-report}
\end{figure}

\begin{figure}[!t]
\centering
\includegraphics[width=1\columnwidth]{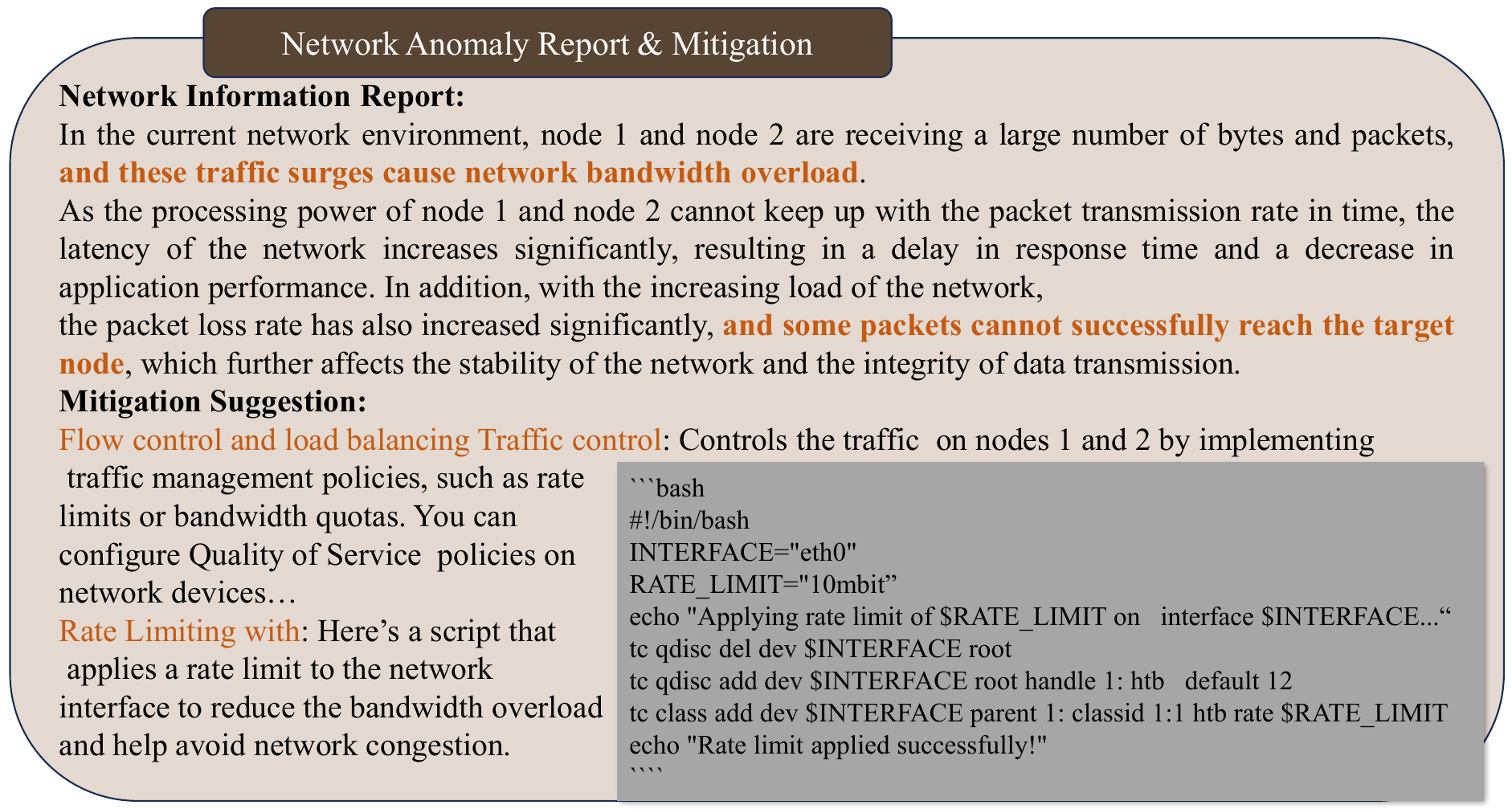}
\caption{Network anomaly report \& mitigation. }
\label{llm-report}
\end{figure}

\begin{table}[!t]
\caption{ROUGE Evaluation of Different LLM With Network Semanticization} 
\label{llm-perforcom}  
\centering  
\begin{tabular}{l *{2}{c} cc}
\hline  
\multirow{2}{*}{Method} & \multicolumn{2}{c}{ROUGE-1} & \multirow{2}{*}{Expert Score $\uparrow$} & \multirow{2}{*}{BLEU $\uparrow$} \\  
\cline{2-3}
& Recall $\uparrow$ & Precision $\uparrow$  & & \\  
\hline  
Llama3 & 34.47 & 26.30  & 83.01 & 26.5 \\   
GPT-3.5 & 22.81 & 26.70  & 82.51 & 22.3 \\   
GPT-4 & 42.23 & 32.58& 85.20 & 35.8 \\  
DeepSeek & \textbf{54.76} & \textbf{45.34}  & \textbf{86.88} & \textbf{48.6} \\  
\hline  
\end{tabular}
\end{table}

\subsection{Text Quality Assessment Experiment}

To answer Question 3, we designed an experiment to evaluate MSADM against other generative text baseline models. This assessment aims to examine the quality of MSADM's generated content and determine whether it meets the requirements for common LLM tasks.\par

Comparative evaluations against state-of-the-art text generation models in Table~\ref{gentext-perforcom} demonstrate MSADM's significant advantages, achieving 35.63 ROUGE-1 F1 (127.8\% higher than KEGNN's 15.64) and 13.87 ROUGE-2 F1 (381.6\% improvement over CTR's 2.14). These quantitative improvements confirm our model's dual capability in eliminating data redundancy while preserving mission-critical network semantics, primarily enabled by its context-aware feature distillation mechanism and domain-specific linguistic structuring paradigm that adaptively prioritizes operational maintenance requirements.\par

Based on the ROUGE evaluation results shown in Table~\ref{llm-perforcom}, the proposed MSADM method demonstrates robust performance when integrated with various LLM. Among them, the DeepSeek model exhibits the most outstanding performance across all metrics, achieving a ROUGE-1 recall of 54.76\%, precision of 45.34\%, expert score of 86.88, significantly outperforming other comparison models. GPT-4 ranked second in expert score (85.20) and ROUGE-1 recall (42.23), while Llama3 and GPT-3.5 demonstrated advantages in specific metrics. These results confirm the MSADM method's strong model compatibility and adaptability.

\subsection{LLM Reports for Network Maintenance Guidance}

To validate operational maintenance capabilities (Q4), we simulate network congestion caused by misconfigured traffic surges (100MB to 200MB).\par

As shown in Fig.~\ref{msadm-report}, MSADM's semantic rule tree identified critical anomalies: node-level packet loss (44.0\%) and bit error rates (22.0\%) exhibited sharp nonmonotonic fluctuations, while link-layer analysis revealed sustained transmission delay escalation and malicious traffic patterns. These structured semantics quantified bandwidth overload impacts on Nodes 1-2, where abrupt traffic growth overwhelmed processing capacity, leading to application performance degradation and packet delivery failures.\par

The LLM-generated report (Fig.~\ref{llm-report}) diagnosed the root cause as a mismatch between traffic volume and node processing capabilities, proposing executable rate-limiting solutions. The provided \texttt{tc} script enforces 10mbit bandwidth caps on \texttt{eth0} interfaces through traffic control policies, demonstrating how semanticized network states enable precise mitigation strategies without predefined templates. \par

\section{Discussion}
We have illustrated the advantages of our scheme for assisting network operators with health management in HNs. In this section, we explore potential future directions in conjunction with our scheme.

To better adapt to the diverse communicating entities in the HNs, we deliberately made trade-offs to enhance the model's scalability. Currently, we model KPIs commonly owned by each entity. However, this approach overlooks the intricate interactions between higher layers, such as the transport protocols they utilize, network layer TM mechanisms, and potential device interactions. A promising future direction involves leveraging MSADM to model the state behavior of higher-level network participants (e.g., Web Server,~SQL Server), such as the application layer, and integrating them with our scheme to form a network for microservice architecture-based anomaly detection solutions.\par

Current LLMs face challenges of high computational demands and resource consumption. Constrained by device performance, they struggle to achieve deployment and application within networks. Fortunately, future approaches will explore lighter-weight LLM designs and segmented inference techniques to address these issues. For instance, research by Lü Ze et al.~\cite{lyu2025larger} proposes a computation-distributed solution that maintains accuracy while enabling rapid inference by reducing latency and bandwidth consumption.

\section{Conclusion and Future Work}
We introduced semantic representation in wireless networks and developed an end-to-end health management scheme for HNs aided by LLM. Our scheme automatically processed the collected anomaly data, predicted anomaly categories, and provided mitigation options. To address the inability of basic rule-based or machine-learning-based systems to adapt to multi-device environments, we proposed MSADM. MSADM monitored the status of entity communication KPIs using a predefined rule base, performed anomaly detection and classification via a rule-enhanced converter structure, and generated uniform and scaled textual representations of anomalies using a semantic rule tree. In addition, the inclusion of chain-of-thinking-based LLM in the diagnosis process enhanced fault diagnosis and generated detailed reports to locate faults and recommend optimization strategies accurately. Experiments demonstrated that the accuracy of MSADM's anomaly detection exceeded that of current mainstream models. Additionally, the experimentally generated anomaly reports and solutions illustrated the potential of our approach in improving the efficiency and accuracy of intelligent O\&M analysis of networks.

\bibliographystyle{IEEEtran}
\bibliography{reference}

@book{ChineseAcademyofCyberspaceStudies2024,
author = {Chinese Academy of Cyberspace Studies},
title = {World Internet Development Report 2023},
year = {2024},
publisher = {Sciendo}
}

@ARTICLE{11244851,
  author={Yuan, Xun and Wang, Xiaonan and Kato, Mina and Tang, Fengxiao and Li, Yangfan and Zhao, Ming and Kato, Nei},
  journal={IEEE Transactions on Networking}, 
  title={CiiNet: Self-Iterative Performance Optimization for Dynamic Networks Based on Causal Inference and Interpretable Evaluation}, 
  year={2025},
  volume={},
  number={},
  pages={1-15},
  doi={10.1109/TON.2025.3626551}}

@ARTICLE{THz,
  author={Yuan, Xun and Tang, Fengxiao and Zhao, Ming and Kato, Nei},
  journal={IEEE Transactions on Wireless Communications}, 
  title={Joint Rate and Coverage Optimization for the THz/RF Multi-band Communications of Space-air-ground Integrated Network in 6G}, 
  year={2023},
}

@article{tuli2022tranad,
  title={Tranad: Deep transformer networks for anomaly detection in multivariate time series data},
  author={Tuli, Shreshth and Casale, Giuliano and Jennings, Nicholas R},
  journal={arXiv preprint arXiv:2201.07284},
  year={2022}
}

@article{lyu2025larger,
  title={The larger the merrier? Efficient large AI model inference in wireless edge networks},
  author={Lyu, Zhonghao and Xiao, Ming and Xu, Jie and Skoglund, Mikael and Di Renzo, Marco},
  journal={arXiv preprint arXiv:2505.09214},
  year={2025}
}

@misc{wu2024nextgptanytoanymultimodalllm,
      title={NExT-GPT: Any-to-Any Multimodal LLM}, 
      author={Shengqiong Wu and Hao Fei and Leigang Qu and Wei Ji and Tat-Seng Chua},
      year={2024},
      eprint={2309.05519},
      archivePrefix={arXiv},
      primaryClass={cs.AI},
}

@article{Fu2022ComplexityBasedPF,
  title={Complexity-Based Prompting for Multi-Step Reasoning},
  author={Yao Fu and Hao-Chun Peng and Ashish Sabharwal and Peter Clark and Tushar Khot},
  journal={ArXiv},
  year={2022},
  volume={abs/2210.00720},
  url={https://api.semanticscholar.org/CorpusID:252683303}
}

@article{hayat2016survey,
  title={Survey on unmanned aerial vehicle networks for civil applications: A communications viewpoint},
  author={Hayat, Samira and Yanmaz, Ev{\c{s}}en and Muzaffar, Raheeb},
  journal={IEEE Communications Surveys \& Tutorials},
  year={2016},
  publisher={IEEE}
}

@article{boem2019distributed,
  title={Distributed fault-tolerant control of large-scale systems: An active fault diagnosis approach},
  author={Boem, Francesca and Gallo, Alexander J and Raimondo, Davide M and Parisini, Thomas},
  journal={IEEE Transactions on Control of Network Systems},
  year={2019},
  publisher={IEEE}
}

@article{szilagyi2012automatic,
  title={An automatic detection and diagnosis framework for mobile communication systems},
  author={Szil{\'a}gyi, P{\'e}ter and Nov{\'a}czki, Szabolcs},
  journal={IEEE transactions on Network and Service Management},
  year={2012},
  publisher={IEEE}
}

@article{barco2008continuous,
  title={Continuous versus discrete model in autodiagnosis systems for wireless networks},
  author={Barco, Raquel and L{\'a}zaro, Pedro and D{\'\i}ez, Luis and Wille, Volker},
  journal={IEEE Transactions on Mobile Computing},
  year={2008},
  publisher={IEEE}
}

@article{khanafer2008automated,
  title={Automated diagnosis for UMTS networks using Bayesian network approach},
  author={Khanafer, Rana M and Solana, Beatriz and Triola, Jordi and Barco, Raquel and Moltsen, Lars and Altman, Zwi and Lazaro, Pedro},
  journal={IEEE Transactions on vehicular technology},
  year={2008},
  publisher={IEEE}
}

@article{barco2010learning,
  title={Learning of model parameters for fault diagnosis in wireless networks},
  author={Barco, Raquel and Wille, Volker and D{\'\i}ez, Luis and Toril, Mat{\'\i}as},
  journal={Wireless Networks},
  year={2010},
  publisher={Springer}
}

@article{chen2024aigc,
  title={AIGC-based evolvable digital twin networks: A road to the intelligent metaverse},
  author={Chen, Xuehan and Luo, Linfeng and Tang, Fengxiao and Zhao, Ming and Kato, Nei},
  journal={IEEE Network},
  volume={38},
  number={6},
  pages={370--379},
  year={2024},
  publisher={IEEE}
}

@article{khatib2015diagnosis,
  title={Diagnosis based on genetic fuzzy algorithms for LTE self-healing},
  author={Khatib, Emil J and Barco, Raquel and G{\'o}mez-Andrades, Ana and Serrano, Inmaculada},
  journal={IEEE Transactions on vehicular technology},
  year={2015},
  publisher={IEEE}
}

@article{jin2015detecting,
  title={DETECTING NODE FAILURES IN MOBILE WIRELESS NETWORKS: A PROBABILISTIC APPROACHetecting node failures in mobile wireless networks: a probabilistic approach},
  author={KarthikeyJini, MRuofan and VWanitha Devi, S and Srinivasan, J and Arulpg, Bing and Wei, Wei and Zhang, Xiaolan and Chen, Xian and Bar-Shalom, Yaakov and Willett, Peter},
  journal={IEEE Transath, Actions on Mobile Computing},
  year={2015},
  publisher={IEEE}
}

@article{qian2021detection,
  title={Detection of mobile network abnormality using deep learning models on massive network measurement data},
  author={Qian, Bing and Lu, Shun},
  journal={Computer Networks},
  year={2021},
  publisher={Elsevier}
}

@article{nti2022mini,
  title={A mini-review of machine learning in big data analytics: Applications, challenges, and prospects},
  author={Nti, Isaac Kofi and Quarcoo, Juanita Ahia and Aning, Justice and Fosu, Godfred Kusi},
  journal={Big Data Mining and Analytics},
  year={2022},
  publisher={TUP}
}

@article{chen2023imdiffusion,
  title={Imdiffusion: Imputed diffusion models for multivariate time series anomaly detection},
  author={Chen, Yuhang and Zhang, Chaoyun and Ma, Minghua and Liu, Yudong and Ding, Ruomeng and Li, Bowen and He, Shilin and Rajmohan, Saravan and Lin, Qingwei and Zhang, Dongmei},
  journal={arXiv preprint arXiv:2307.00754},
  year={2023}
}

@ARTICLE{10806770,
  author={Tang, Fengxiao and Luo, Linfeng and Guo, Zhiqi and Li, Yangfan and Zhao, Ming and Kato, Nei},
  journal={IEEE Transactions on Mobile Computing}, 
  title={Semi-Distributed Network Fault Diagnosis Based on Digital Twin Network in Highly Dynamic Heterogeneous Networks}, 
  year={2024},
  volume={},
  number={},
  pages={1-14},
  doi={10.1109/TMC.2024.3519576}}

@inproceedings{yang2023dcdetector,
  title={Dcdetector: Dual attention contrastive representation learning for time series anomaly detection},
  author={Yang, Yiyuan and Zhang, Chaoli and Zhou, Tian and Wen, Qingsong and Sun, Liang},
  booktitle={Proceedings of the 29th ACM SIGKDD Conference on Knowledge Discovery and Data Mining},
  year={2023}
}

@inproceedings{tariq2019detecting,
  title={Detecting anomalies in space using multivariate convolutional LSTM with mixtures of probabilistic PCA},
  author={Tariq, Shahroz and Lee, Sangyup and Shin, Youjin and Lee, Myeong Shin and Jung, Okchul and Chung, Daewon and Woo, Simon S},
  booktitle={Proceedings of the 25th ACM SIGKDD international conference on knowledge discovery \& data mining},
  year={2019}
}

@inproceedings{ren2019time,
  title={Time-series anomaly detection service at microsoft},
  author={Ren, Hansheng and Xu, Bixiong and Wang, Yujing and Yi, Chao and Huang, Congrui and Kou, Xiaoyu and Xing, Tony and Yang, Mao and Tong, Jie and Zhang, Qi},
  booktitle={Proceedings of the 25th ACM SIGKDD international conference on knowledge discovery \& data mining},
  year={2019}
}

@ARTICLE{10113753,
  author={Zhao, Haihong and Yang, Bo and Cui, Jiaxu and Xing, Qianli and Shen, Jiaxing and Zhu, Fujin and Cao, Jiannong},
  journal={IEEE Transactions on Mobile Computing}, 
  title={Effective Fault Scenario Identification for Communication Networks via Knowledge-Enhanced Graph Neural Networks}, 
  year={2024},
  volume={23},
  number={4},
  pages={3243-3258},
  doi={10.1109/TMC.2023.3271715}}

@article{jeong2023anomalybert,
  title={Anomalybert: Self-supervised transformer for time series anomaly detection using data degradation scheme},
  author={Jeong, Yungi and Yang, Eunseok and Ryu, Jung Hyun and Park, Imseong and Kang, Myungjoo},
  journal={arXiv preprint arXiv:2305.04468},
  year={2023}
}

@article{lyu2022knowledge,
  title={Knowledge enhanced graph neural networks for explainable recommendation},
  author={Lyu, Ziyu and Wu, Yue and Lai, Junjie and Yang, Min and Li, Chengming and Zhou, Wei},
  journal={IEEE Transactions on Knowledge and Data Engineering},
  year={2022},
  publisher={IEEE}
}

@inproceedings{zou2024fta,
  title={FTA-detector: Troubleshooting Gray Link Failures Based on Fault Tree Analysis},
  author={Zou, Yan and Pan, Tian and Fu, Qiang and Jia, Chenhao and Yi, Qingqiang and Wan, Ying and Zhang, Jiao and Huang, Tao},
  booktitle={NOMS 2024-2024 IEEE Network Operations and Management Symposium},
  year={2024},
  organization={IEEE}
}

@inproceedings{10.1145/3077136.3080822,
author = {Li, Piji and Wang, Zihao and Ren, Zhaochun and Bing, Lidong and Lam, Wai},
title = {Neural Rating Regression with Abstractive Tips Generation for Recommendation},
year = {2017},
booktitle = {Proceedings of the 40th International ACM SIGIR Conference on Research and Development in Information Retrieval}
}

@inproceedings{wang2011collaborative,
  title={Collaborative topic modeling for recommending scientific articles},
  author={Wang, Chong and Blei, David M},
  booktitle={Proceedings of the 17th ACM SIGKDD international conference on Knowledge discovery and data mining},
  year={2011}
}

@inproceedings{chen2019co,
  title={Co-attentive multi-task learning for explainable recommendation.},
  author={Chen, Zhongxia and Wang, Xiting and Xie, Xing and Wu, Tong and Bu, Guoqing and Wang, Yining and Chen, Enhong},
  booktitle={IJCAI},
  year={2019}
}

@inproceedings{tang2023swinlstm,
  title={Swinlstm: Improving spatiotemporal prediction accuracy using swin transformer and lstm},
  author={Tang, Song and Li, Chuang and Zhang, Pu and Tang, RongNian},
  booktitle={Proceedings of the IEEE/CVF International Conference on Computer Vision},
  year={2023}
}

@INPROCEEDINGS{10515433,
  author={R, Selvam and S, Velliangiri},
  booktitle={2024 International Conference on Distributed Computing and Optimization Techniques (ICDCOT)}, 
  title={An Improving Intrusion Detection Model Based on Novel CNN Technique Using Recent CIC-IDS Datasets}, 
  year={2024},
  volume={},
  number={},
  pages={1-6},
  doi={10.1109/ICDCOT61034.2024.10515433}}

@article{serdaly2024improving,
  title={IMPROVING NETWORK INTRUSION DETECTION USING THE MINI-VGGNET ARCHITECTURE: TACKLING CHALLENGES OF IMBALANCED DATA.},
  author={Serdaly, Altynay and Imanbekkyzy, Dana and Akhmetay, Adil and Omarov, Batyrkhan},
  journal={Journal of Problems in Computer Science and Information Technologies},
  volume={2},
  number={4},
  pages={3--16},
  year={2024}
}

@article{zhao2025research,
  title={Research on intrusion detection model based on improved MLP algorithm},
  author={Zhao, Qihao and Wang, Fuwei and Wang, Weimin and Zhang, Tianxin and Wu, Haodong and Ning, Weijun},
  journal={Scientific reports},
  volume={15},
  number={1},
  pages={5159},
  year={2025},
  publisher={Nature Publishing Group UK London}
}

@Article{technologies13030088,
AUTHOR = {Imani, Mehdi and Beikmohammadi, Ali and Arabnia, Hamid Reza},
TITLE = {Comprehensive Analysis of Random Forest and XGBoost Performance with SMOTE, ADASYN, and GNUS Under Varying Imbalance Levels},
JOURNAL = {Technologies},
VOLUME = {13},
YEAR = {2025},
NUMBER = {3},
ARTICLE-NUMBER = {88},
ISSN = {2227-7080},
DOI = {10.3390/technologies13030088}
}

@ARTICLE{10644103,
  author={Kato, Mina and Koketsu Rodrigues, Tiago and Abe, Toru and Suganuma, Takuo},
  journal={IEEE Internet of Things Journal}, 
  title={Exploiting Radio Frequency Characteristics With a Support Unmanned Aerial Vehicle to Improve Wireless Sensor Location Estimation Accuracy}, 
  year={2024},
  volume={11},
  number={24},
  pages={39570-39578},
  doi={10.1109/JIOT.2024.3448394}}

@ARTICLE{10981691,
  author={Yuan, Xun and Wang, Xiaonan and Tang, Fengxiao and Zhou, Qingping and Zhao, Ming and Kato, Nei},
  journal={IEEE Transactions on Networking}, 
  title={MPITE: Multidimensional Performance Evaluator for Interpretable and Traceable Network Performance Evaluation}, 
  year={2025},
  volume={},
  number={},
  pages={1-16},
  doi={10.1109/TON.2025.3562348}}

@article{10.1016/j.engappai.2004.08.031,
author = {Neema, Sandeep and Bapty, Ted and Shetty, Shweta and Nordstrom, Steven},
title = {Autonomic fault mitigation in embedded systems},
year = {2004},
issue_date = {October, 2004},
publisher = {Pergamon Press, Inc.},
address = {USA},
volume = {17},
number = {7},
issn = {0952-1976},
doi = {10.1016/j.engappai.2004.08.031},
journal = {Eng. Appl. Artif. Intell.},
month = oct,
pages = {711–725},
numpages = {15}
}

@ARTICLE{10591707,
  author={He, Ying and Fang, Jingcheng and Yu, F. Richard and Leung, Victor C.},
  journal={IEEE Transactions on Mobile Computing}, 
  title={Large Language Models (LLMs) Inference Offloading and Resource Allocation in Cloud-Edge Computing: An Active Inference Approach}, 
  year={2024},
  volume={23},
  number={12},
  pages={11253-11264},
  doi={10.1109/TMC.2024.3415661}}

@ARTICLE{9130088,
  author={Wang, Zhaorui and Liu, Liang and Cui, Shuguang},
  journal={IEEE Transactions on Wireless Communications}, 
  title={Channel Estimation for Intelligent Reflecting Surface Assisted Multiuser Communications: Framework, Algorithms, and Analysis}, 
  year={2020},
  volume={19},
  number={10},
  pages={6607-6620},
  doi={10.1109/TWC.2020.3004330}}

@inproceedings{10.1145/3651890.3672268,
author = {Wu, Duo and Wang, Xianda and Qiao, Yaqi and Wang, Zhi and Jiang, Junchen and Cui, Shuguang and Wang, Fangxin},
title = {NetLLM: Adapting Large Language Models for Networking},
year = {2024},
isbn = {9798400706141},
publisher = {Association for Computing Machinery},
address = {New York, NY, USA},
url = {https://doi.org/10.1145/3651890.3672268},
doi = {10.1145/3651890.3672268},
booktitle = {Proceedings of the ACM SIGCOMM 2024 Conference},
pages = {661–678},
numpages = {18},
location = {Sydney, NSW, Australia},
series = {ACM SIGCOMM '24}
}

\begin{IEEEbiography}[{\includegraphics[width=1in,height=1.25in,clip,keepaspectratio]{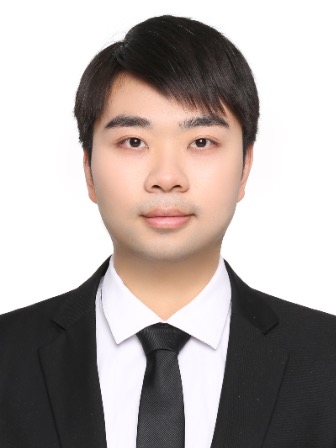}}]
\footnotesize \textbf{Fengxiao Tang (Senior Member, IEEE)}
 is a full professor with the School of Computer Science and Engineering, Central South University, and a distinguished professor in the Graduate School of Information Sciences(GSIS), Tohoku University. He has been an Assistant Professor from 2019 to 2020 and an Associate Professor from 2020 to 2021 at the Graduate School of Information Sciences (GSIS) of Tohoku University. His research interests are unmanned aerial vehicle systems, IoT security, game theory optimization, network traffic control, and machine learning algorithms. He was a recipient of the prestigious Dean's and President's Awards from Tohoku University in 2019, and several best paper awards at conferences, including IC-NIDC 2018/2023, GLOBECOM 2017/2018. He was also a recipient of the prestigious Funai Research Award in 2020, the IEEE ComSoc Asia-Pacific (AP) Outstanding Paper Award in 2020, and the IEEE ComSoc AP Outstanding Young Researcher Award in 2021.
\end{IEEEbiography}

\begin{IEEEbiography}[{\includegraphics[width=1in,height=1.25in,clip,keepaspectratio]{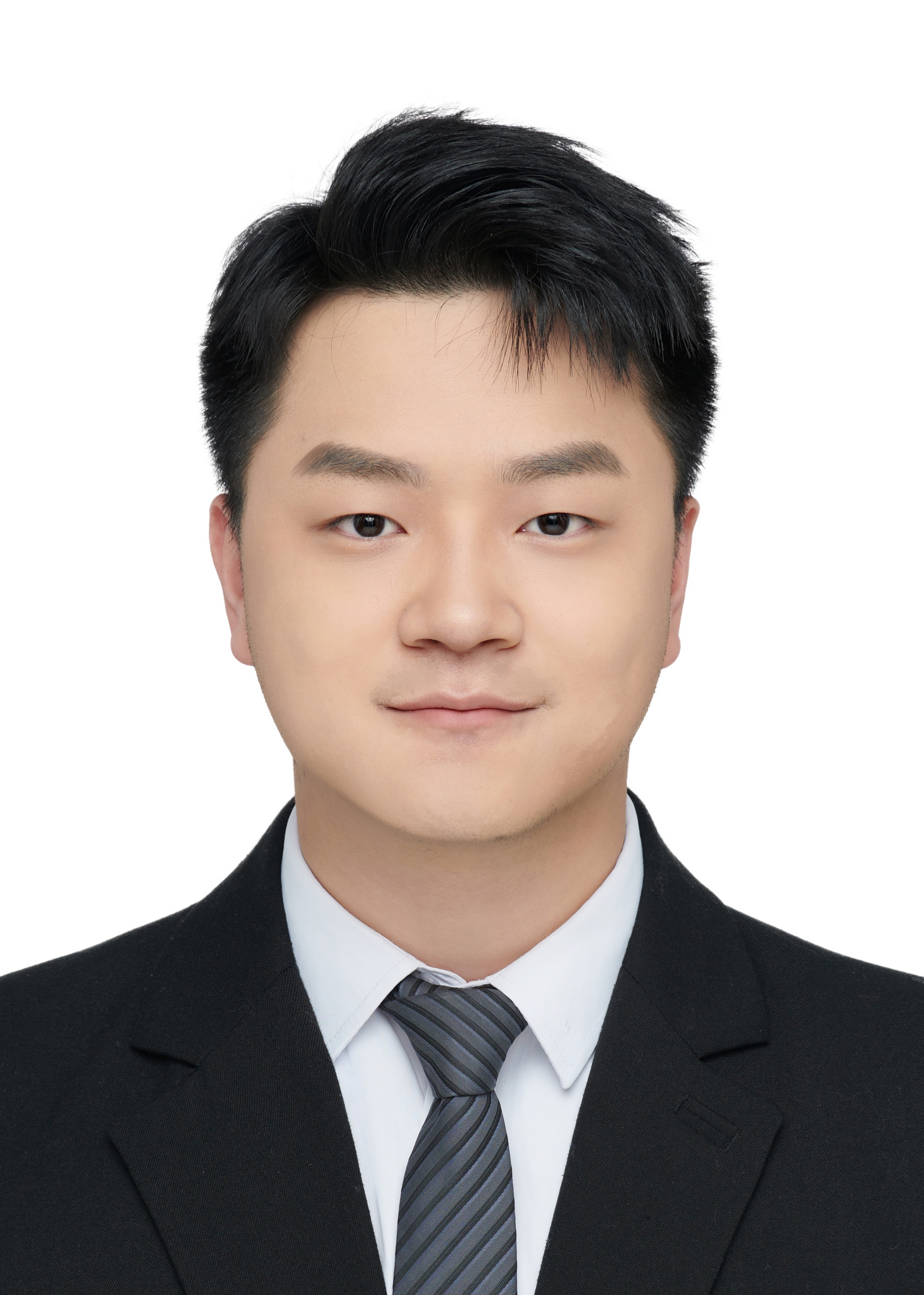}}]
\footnotesize \textbf{Xiaonan Wang} received his Bachelor's degree in Software Engineering from Zhongyuan Institute of Technology, China, in 2022. He has been awarded a Master's degree in Software Engineering at Xinjiang University through joint training.  And he is pursuing a Ph.D. degree at Central South University. His research interests focus on network fault diagnosis and communication network health management.
\end{IEEEbiography}

\begin{IEEEbiography}[{\includegraphics[width=1in,height=1.25in,clip,keepaspectratio]{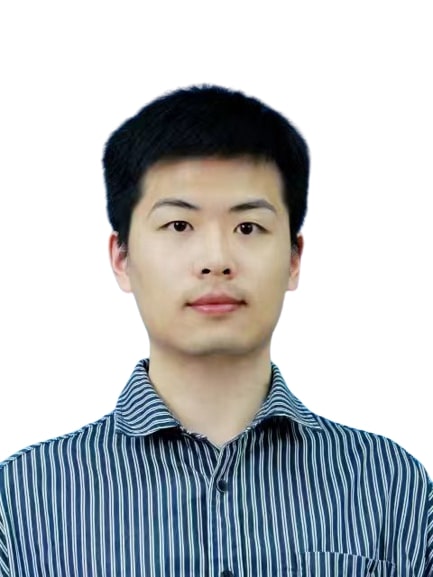}}]
\footnotesize \textbf{Xun Yuan} received his B.E. degree in Software Engineering from Huazhong University of Science and Technology in 2019, and his M.S. degree in Computer Technology from the same university in 2022. He is currently pursuing a Ph.D. degree in the Intelligent Networks and Collaborative Computing Laboratory at Central South University. He is currently undergoing a two-year joint Ph.D. program as a special research student at the Graduate School of Information Science, Tohoku University, Japan. 
\end{IEEEbiography}

\begin{IEEEbiography}[{\includegraphics[width=1in,height=1.25in,clip,keepaspectratio]{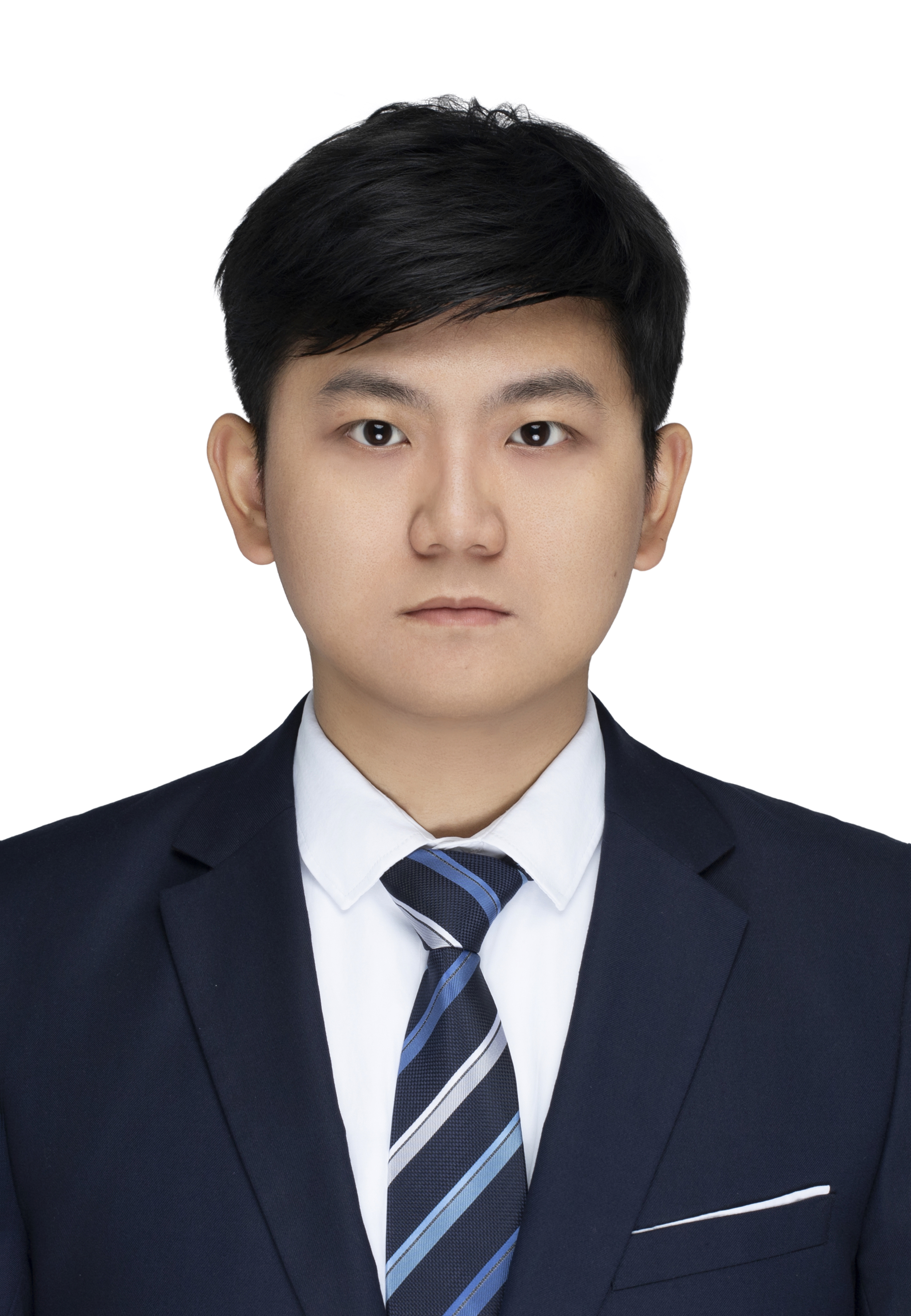}}]
\footnotesize \textbf{LinFeng Luo}
received the B.S. degree in Computer Science and Technology from Huazhong University of Science and Technology, Wuhan, China, in 2019, and the M.S. degree in Software Engineering from the School of Computer Science and Engineering, Central South University, Changsha, China, in 2024. He is currently pursuing a PhD degree in computer science and technology at Central South University, with research interests including Digital Twin Networks, Network Fault Diagnosis, and Graph Neural Networks.
\end{IEEEbiography}

\begin{IEEEbiography}[{\includegraphics[width=1in,height=1.25in,clip,keepaspectratio]{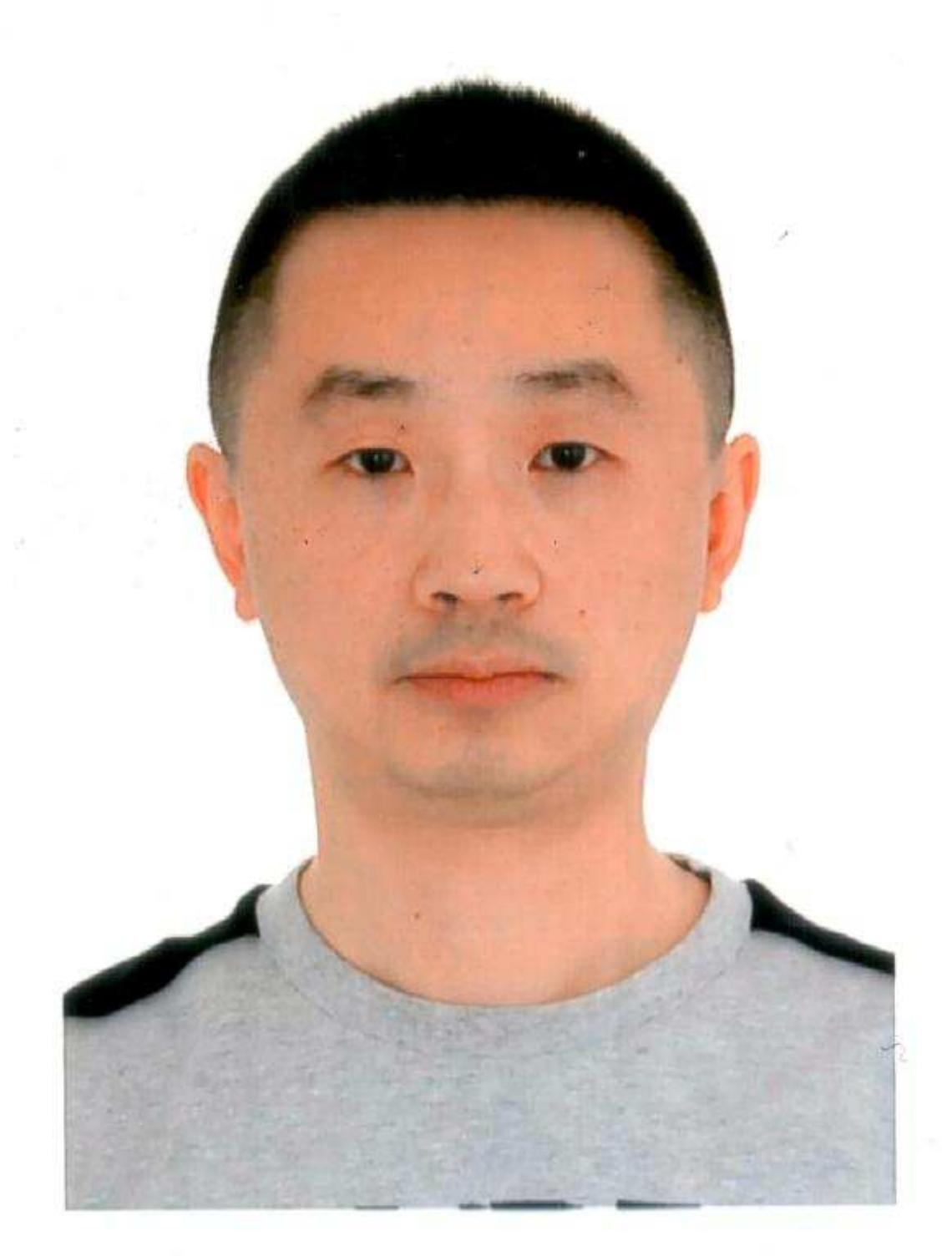}}]
\footnotesize \textbf{Ming Zhao}
received the M.Sc. and Ph.D. degrees in computer science from Central South University, Changsha, China, in 2003 and 2007, respectively. He is currently a Professor at the School of Computer Science and Engineering, Central South University. His main research focuses on wireless networks. He is also a Member of the China Computer Federation.
\end{IEEEbiography}


\begin{IEEEbiography}[{\includegraphics[width=1in,height=1.25in,clip,keepaspectratio]{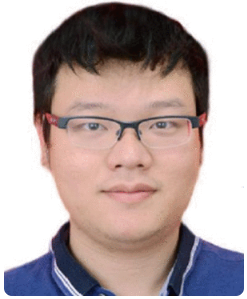}}]
\footnotesize \textbf{Tianchi Huang}
received the PhD degree from the Department of Computer Science and Technology, Tsinghua University, in 2023. His research work focuses on multimedia network streaming, including transmitting streams and edge-assisted content delivery. He has been the reviewer for IEEE Transactions on Vehicular Technology, IEEE Transactions on Mobile Computing, and IEEE Transactions on Multimedia. He received the Best Student Paper Award presented by the ACM Multimedia System 2019 Workshop and the Best Paper Nomination presented by ACM Multimedia Asia 2022. He currently works in the R\&D Department. Sony.
\end{IEEEbiography}

\vspace{-2\baselineskip}

\begin{IEEEbiography}[{\includegraphics[width=1in,height=1.25in,clip,keepaspectratio]{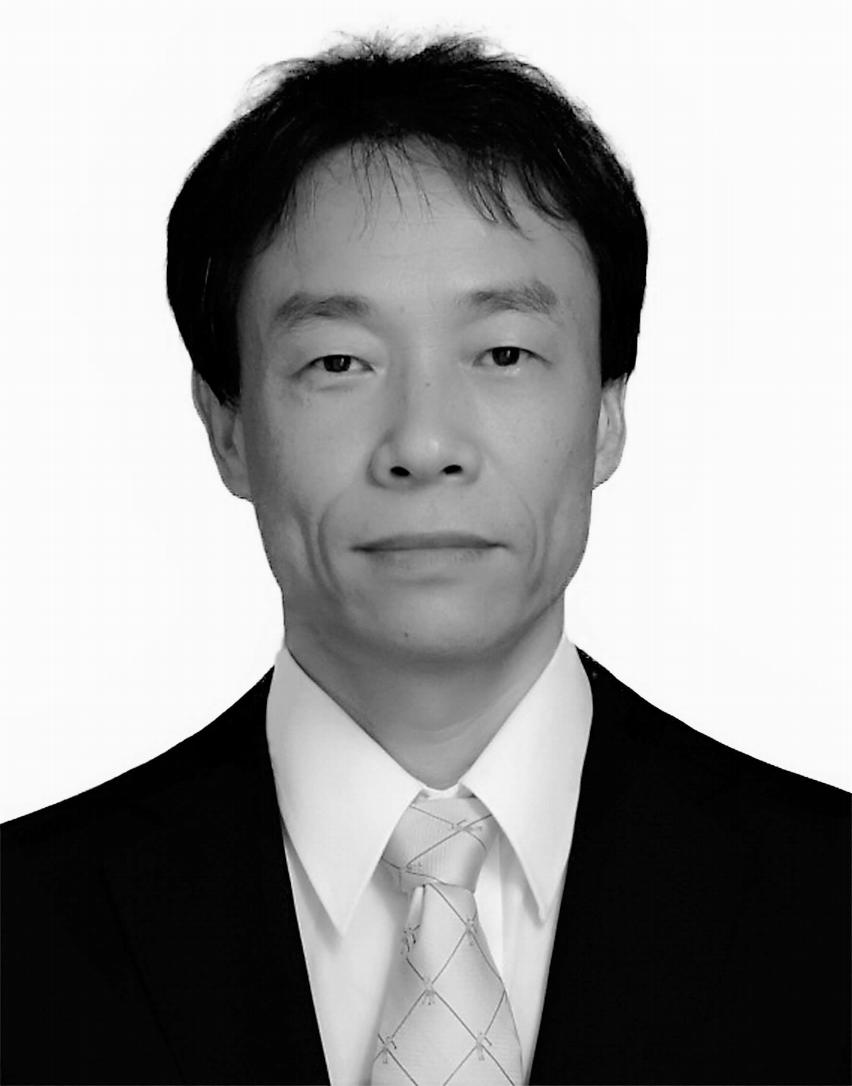}}]
\footnotesize \textbf{Nei Kato (M'04-SM'05-F'13)}
is a full professor and the Dean of the Graduate School of Information Sciences, Tohoku University. He has been engaged in research on computer networking, wireless mobile communications, satellite communications, ad hoc and sensor and mesh networks, smart grid, AI, IoT, big data, and pattern recognition. He was the Vice-President (Member \& Global Activities) of IEEE Communications Society (2018-2019), the Editor-in-Chief of IEEE Transactions on Vehicular Technology (2017-2020), and the Editor-in-Chief of IEEE Network (2015-2017). 
\end{IEEEbiography}

\end{document}